\documentclass[twocolumn,times]{aastex631}
\usepackage{amsmath} 
\usepackage{nicefrac}

\shorttitle{Equatorial Waves and Superrotation on Titan}
\shortauthors{}

\begin{document}

\title{\vspace*{-.2in} \large Equatorial Waves and Superrotation in the Stratosphere of a Titan General Circulation Model}

\author{Neil T. Lewis}
\affiliation{Atmospheric, Oceanic and Planetary Physics, University of Oxford, Oxford, UK}

\affiliation{Department of Mathematics and Statistics, University of Exeter, Exeter, UK}

\author{Nicholas A. Lombardo}
\affiliation{Department of Earth and Planetary Sciences, Yale University, New Haven, CT, USA}

\author{Peter L. Read}
\affiliation{Atmospheric, Oceanic and Planetary Physics, University of Oxford, Oxford, UK}

\author{Juan M. Lora}
\affiliation{Department of Earth and Planetary Sciences, Yale University, New Haven, CT, USA}

\correspondingauthor{N.~T. Lewis}
\email{n.t.lewis@exeter.ac.uk}

\begin{abstract}
\noindent We investigate the characteristics of equatorial waves associated with the maintenance of superrotation in the stratosphere of a Titan general circulation model. A variety of equatorial waves are present in the model atmosphere, including equatorial Kelvin waves, equatorial Rossby waves, and mixed Rossby--gravity waves. In the upper stratosphere, acceleration of superrotation is strongest around solstice and is due to interaction between equatorial Kelvin waves and Rossby-type waves in winter-hemisphere mid-latitudes. The existence of this `Rossby--Kelvin'-type wave appears to depend on strong meridional shear of the background zonal wind that occurs in the upper stratosphere at times away from the equinoxes. In the lower stratosphere, acceleration of superrotation occurs throughout the year and is partially induced by equatorial Rossby waves, which we speculate are generated by quasigeostrophic barotropic instability.  Acceleration of superrotation is generally due to waves with phase speeds close to the zonal velocity of the mean flow. Consequently, they have short vertical wavelengths which are close to the model's vertical grid scale, and therefore are likely to be not properly represented. We suggest this may be a common issue amongst Titan GCMs which should be addressed by future model development. 
\end{abstract}

\vspace*{-.5in}

\keywords{Titan (2186); Planetary Atmospheres (1244); Planetary Science (1255) \vspace*{.3in}}

\section{Introduction} \label{sec:intro}

\subsection{Background}

Titan's stratospheric circulation is dominated by a broad prograde zonal jet, which extends from pole to pole and reaches speeds in excess of $200\,\text{m\,s}^{-1}$ \citep{2008Icar..194..263A,2021Icar..35414030S}. The structure of this jet varies with Titan's seasonal cycle, with maximum zonal velocities obtained in mid-autumn \citep{2021Icar..35414030S}. A property of Titan's zonal jet is that it is \emph{superrotating}, which means that it has more axial angular momentum, $m$, than the solid planet at the equator \citep{1986QJRMS.112..253R,2018AREPS..46..175R}. This can be expressed mathematically by defining a `local superrotation index' \citep{1986QJRMS.112..253R}\begin{equation}
    s\equiv\frac{m}{\Omega a^{2}}-1.
\end{equation}
Above, $\Omega$ is the planetary rotation rate, $a$ is the planetary radius, and $m$ is defined $m=a\cos\vartheta(\Omega a\cos\vartheta +u)$, where $\vartheta$ is latitude and $u$ is the zonal velocity. An atmosphere is then superrotating at any location where $s>0$. A maximum of $m$ located away from the equator will typically be inertially unstable \citep{1981JAtS...38.2354D}, and thus $s>0$ anywhere generally implies the existence of a prograde zonal flow at the equator (although not necessarily a zonal wind maximum), as is the case on Titan. In the presence of friction and dissipative processes which mix $m$ down gradient, non-axisymmetric eddy motions are required to generate $s>0$ \citep{1969JAtS...26..841H,1975JAtS...32.1038G,1979JAtS...36..377R,1986QJRMS.112..253R}. 

In an effort to understand and interpret Titan's observed atmospheric circulation, a number of general circulation models (GCMs) of Titan's atmosphere have been developed. They include: the K\"{o}ln Titan model \citep{1999P&SS...47..493T}, TitanCAM \citep{2009P&SS...57.1931F}, TitanWRF \citep{2011Icar..213..636N}, the IPSL Titan model \citep{2012Icar..218..707L}, and the Titan Atmospheric Model \citep[TAM;][]{2015Icar..250..516L}. Each of these models integrates the hydrostatic primitive equations forwards in time to simulate Titan's atmospheric circulation, forced by radiative heating designed to imitate that in Titan's real atmosphere. A major success of TitanWRF, the IPSL model, and TAM has been reproducing Titan's stratospheric superrotation with a strength similar to that inferred from Cassini observations \citep{2011Icar..213..636N,2012Icar..218..707L,2015Icar..250..516L}. Strong superrotation is not generated in TitanCAM or the K\"{o}ln model, possibly due to poor conservation of angular momentum in their dynamical cores \citep{2012JGRE..11712004L}. 

TitanWRF, the IPSL model, and TAM present us with an opportunity to investigate the properties of wave-like disturbances that exist under Titan-like atmospheric conditions, and lead to the acceleration of superrotation. To this end, \citet{2011Icar..213..636N} and \citet{2012Icar..218..707L} present analyses of the angular momentum budget and waves obtained in their respective models, TitanWRF and the IPSL model. In both cases, eddy acceleration of the superrotating jet is strongest around the solstices, and acts to balance deceleration at the equator due to mean-flow advection of angular momentum into the winter hemisphere. In the TitanWRF model, eddy accelerations appear to be concentrated into short, intense `transfer events'. Both studies identify that the acceleration of superrotation is predominantly due to waves with low zonal wavenumbers, which they suggest are generated by barotropic instability in sub-tropical latitudes. An equivalent analysis has not been performed for TAM. 

Titan is both small and slowly rotating when compared with the Earth ($a\,{=}\,2576\,\text{km}\,{\approx}\,0.4a_{\text{Earth}}$; $\Omega\,{=}\,4.56\times10^{-6}\,\text{s}^{-1}\,{\approx}\,0.06\Omega_{\text{Earth}}$; \citealp{2018AREPS..46..175R}). The generation of superrotation in the atmospheres of small and/or slowly rotating planets has been investigated using simplified `Earth-like' GCMs \citep[e.g.,][]{2010JGRE..11512008M,2014JAtS...71..596P,2014Icar..238...93D,2018QJRMS.144.2537W,2020P&SS..19004976L,2021JAtS...78.1245L}, typically forced by a Newtonian relaxation towards a time-independent Earth-like radiative-convective equilibrium temperature field \citep[as in, e.g.,][]{1994BAMS...75.1825H}, and with all planetary parameters taken to be those of Earth (except for the radius or rotation rate). These studies have shown that equatorial superrotation emerges naturally in this context when either the planetary radius or rotation rate is reduced from Earth's. Stronger superrotation is obtained when the planetary radius is reduced when compared with that due to an equivalent reduction in the planetary rotation rate \citep{2014Icar..238...93D}. This occurs as a reduction of the rotation rate reduces the ratio of the radiative damping timescale to the advective timescale, which in turn induces a stronger overturning circulation that can more effectively communicate the effect of surface friction to the free atmosphere.

In the simplified Earth-like setting, superrotation is generated by a planetary-scale, zonal wavenumber-1 mode comprised of Rossby waves in mid-latitudes coupled with a Kelvin wave at the equator \citep{2010JGRE..11512008M}. \citet{2005JAtS...62.2514I} and \citet{2014GeoRL..41.4118W} have shown that this sort of mode emerges due to a barotropic, ageostropic instability, which occurs when the meridional shear in the background wind is substantial enough that the Rossby component (moving west relative to the mean flow) is doppler-shifted sufficiently by the mid-latitude jet that it can phase-lock with the Kelvin component (moving east relative to the mean flow). { The potential importance of Rossby--Kelvin waves for maintaining superrotation is not limited to idealised Earth-like models; \citet{2022JGRE..12707164T} identify a similar mode in a simulation of Venus' atmospheric circulation which converges momentum towards the equator. }

In some circumstances, this `Rossby--Kelvin instability' has been found to generate superrotation in an idealised model's spin-up phase, but then disappear when the circulation equilibrates \citep{2010JGRE..11512008M,2016JAtS...73.3181D}. This occurs when the accelerated equatorial superrotation is strong enough to remove the meridional shear required for the Kelvin and Rossby components to phase lock. In this scenario, after spin-up, \citet{2016JAtS...73.3181D} show that equatorial Rossby waves, generated by a mixed barotropic--baroclinic instability, can induce the acceleration required to maintain superrotation. 

The idealised modeling studies described in the previous two paragraphs do not consider the effects of a seasonal cycle, thus omitting an important feature of Titan. \citet{2014ApJ...787...23M} show that including a seasonal cycle in an idealised Earth-like model can suppress superrotation, unless the radiative damping time is set to be very long ($100$ times that in the standard, Earth-like set-up). The radiative timescale on Titan is long compared with Earth's due to its larger atmospheric mass and lower temperatures, and \citet{2014ApJ...787...23M} suggest that this may be critical for the development of strong superrotation on Titan.

\subsection{Aim of this work}

In the present study, we present an analysis of equatorialwaves in a simulation of Titan's atmosphere performed with the TAM general circulation model, focusing on the stratosphere. This work complements and extends that of \cite{2011Icar..213..636N} and \citet{2012Icar..218..707L}.

Our first aim is to characterise the different equatorial waves that are supported in our model of Titan's atmosphere, with an emphasis on investigating their three-dimensional structure, as exploration of this topic in previous work has been limited in scope. We will also determine which of the waves contribute most to accelerating Titan's equatorial superrotation.

The results of our analysis are compared with those obtained from other Titan GCMs \citep{2011Icar..213..636N,2012Icar..218..707L}, as well as results from idealised modeling \citep{2010JGRE..11512008M,2014ApJ...787...23M,2016JAtS...73.3181D} and against expectations from linear theory \citep{2005JAtS...62.2514I,2014GeoRL..41.4118W}.

{Throughout this work, we use the nomenclature `equatorial waves' to describe the waves analysed in this study, even though some that we identify have a meridional scale comparable to the planetary scale. We will show that there is good agreement between the meridional scale of the waves we identify and that predicted by the linear theory for waves on the equatorial beta plane, namely the equatorial deformation radius $l_{d}$ (for which the ratio $l_{\text{d}}/a$ depends inversely on radius and rotation rate; \citealp{1966JMeSJ..44...25M,2009RvGeo..47.2003K}). }

\subsection{Outline of paper}

The rest of this paper is organised as follows. In Section \ref{sec:methods} we describe the TAM general circulation model, as well as the diagnostics we use to analyse the waves in our experiment. In Section \ref{sec:basic_circulation} we present the model's zonally averaged circulation and show how it varies with Titan's seasonal cycle. We also analyse the seasonal cycle of eddy angular momentum fluxes. The main results of this study are then presented in Section \ref{sec:autumn}, where we analyse the waves present in the model atmosphere in northern hemisphere autumn, when eddy momentum flux convergence at the equator is strongest. Discussion is presented in Section \ref{sec:discuss}, and our main conclusions are summarised in Section \ref{sec:conclude}. 

\section{Methods}\label{sec:methods}

\subsection{General Circulation Model}

We use results from simulations with the Titan Atmospheric Model GCM. TAM is introduced and described in detail in \citet{2015Icar..250..516L}, and updated in \citet{2017Icar..286..270L}, \citet{2022Icar..37314623B} and \citet{2023Icar..39015291L}. The version of TAM used for this study is identical to the `static' configuration described in \citet{2023Icar..39015291L}, which features an updated radiative transfer code{, and a higher model top to enable a better simulation of the stratosphere.}

The model is built on the GFDL primitive equation spectral dynamical core \citep{1982MWRv..110..625G}. The equations are solved on a spherical domain using a pseudospectral method in the horizontal (prognostic fields represented by a triangular truncation of spherical harmonics), and a finite difference method in the vertical. The vertical coordinate is a hybrid `sigma' coordinate, $\sigma=p/p_{\text{s}}$, that transitions to a pure pressure, $p$, coordinate around $100\,\text{hPa}$. 

The model is run with a T21 triangular truncation, which corresponds to a horizontal resolution of roughly $5.6^{\circ}$ latitude/longitude at the equator. This resolution is relatively low, but necessary to avoid the computational expense of the model becoming prohibitive. { There are 50 layers in the vertical direction up to a top pressure of $1.65\times10^{-4}\,\text{hPa}$ (approximately $500\,\text{km}$ altitude). They are distributed in $\sigma$ according to $\sigma=\exp\left[-16(0.35\tilde{z}+0.65\tilde{z}^{3})\right]$, where $\tilde{z}$ is equally spaced in the unit interval. }
To stabilise the model, an eighth-order horizontal hyperdiffusion acting on a timescale of $3.7\,\text{days}$ is applied throughout the atmosphere. A Laplacian `sponge' is included in the uppermost model layer, and a Rayleigh drag is { applied above $5\times10^{-3}\,\text{hPa}$,} to prevent wave reflection at the model top. The sponge damps eddy zonal and meridional winds on a { timescale of $9\times10^{-3}\,\text{days}$ at the grid scale}. The Rayleigh drag is applied to the full zonal and meridional wind; its strength increases with altitude and acts on a { timescale of $0.2\,\text{days}$} in the uppermost layer. Above, `day' refers to an Earth day ($86400\,\text{s}$). 

TAM is adapted to Titan through the inclusion of a multi-band radiative transfer code with opacities based on the composition of Titan's atmosphere (described in \citealp{2023Icar..39015291L}), a representation of Titan's methane hydrological cycle \citep{2017Icar..286..270L,2022Icar..37314623B}, and an astronomy calculation adapted to give Titan's seasonal cycle of insolation. A Mellor--Yamada scheme is used to parametrise boundary layer turbulence \citep{1982RvGSP..20..851M} and the `Simple Betts--Miller' scheme of \citet{2007JAtS...64.1959F} is used to parametrise convection. For a more detailed description of TAM's physics packages, the reader is directed to \citet{2015Icar..250..516L} and \citet{2023Icar..39015291L}. 

To spin-up the model, the zonal wind is nudged towards a prescribed, hemispherically symmetric superrotating state. Once the total angular momentum of the model's zonal winds reaches that of the prescribed state, the nudging is switched off, and the model allowed to run freely until the circulation reaches a statistically steady state. A detailed description of this procedure is given in \citet{2023Icar..39015291L}. The simulation is then run for a further half Titan year {beginning at northern hemisphere autumn equinox} ({ $14.7$ Earth years; solar longitudes $L_{\text{s}}=180^{\circ}$ -- $360^{\circ}$}) producing output every 3 hours { (where an hour is $3600\,\text{s}$)}, which we interpolate onto a pure pressure coordinate and use for our analysis.

\subsection{Diagnostics}

\subsubsection{Transformed Eulerian Mean momentum budget}

To analyse the acceleration of the mean flow by eddies, we make use of the Transformed Eulerian Mean (TEM) formulation of the zonally-averaged zonal momentum equation. In a log-pressure coordinate, this may be written \citep[][Chapter 3]{1987mad..book.....A} \begin{equation}
    \frac{\text{D}^{\ast}\overline{u}}{\text{D}t} - \frac{\overline{u}\,\overline{v}^{\ast}\tan\vartheta}{a}-f\overline{v}^{\ast} = \frac{1}{\rho_{0} a\cos\vartheta}\nabla\cdot\boldsymbol{F} - \overline{D}, \label{eq:TEM_u}
\end{equation}
where \begin{equation}
    \frac{\text{D}^{\ast}}{\text{D}t} = \frac{\partial}{\partial t} +\frac{\overline{v}^{\ast}}{a}\frac{\partial}{\partial\vartheta} + \overline{w}^{\ast}\frac{\partial}{\partial z^{\ast}}
\end{equation}
is the material derivative following the residual mean meridional circulation ($\overline{v}^{\ast}$,$\overline{w}^{\ast}$) defined by \begin{align}
    \overline{v}^{\ast}&=\overline{v}-\frac{1}{\rho_0}\frac{\partial\rho_0\psi}{\partial z^{\ast}}, \\ 
    \overline{w}^{\ast}&=\overline{w}+\frac{1}{a\cos\vartheta}\frac{\partial\psi\cos\vartheta}{\partial\vartheta},
\end{align}
and $\psi=\overline{v^{\prime}\theta^{\prime}}/(\partial\overline{\theta}/\partial z^{\ast})$. The vector $\boldsymbol{F}=\left(F^{(\vartheta)},F^{(z^{\ast})}\right)$ is the Eliassen--Palm flux (EP flux), defined \begin{align}
    F^{(\vartheta)}&=\rho_0 a\cos\vartheta\left(\frac{\partial\overline{u}}{\partial z^{\ast}}\psi-\overline{v^{\prime}u^{\prime}}\right), \\ 
    F^{(z^{\ast})}&=\rho_0 a\cos\vartheta\left[\left(f-\frac{1}{a\cos\vartheta}\frac{\partial\overline{u}\cos\vartheta}{\partial\vartheta}\right)\psi - \overline{w^{\prime}u^{\prime}}\right], 
\end{align} 
with \begin{equation}
    \nabla\cdot\boldsymbol{F}=\frac{1}{a\cos\vartheta}\frac{\partial F^{(\vartheta)}\cos\vartheta}{\partial\vartheta}+\frac{\partial F^{(z^{\ast})}}{\partial z^{\ast}}. 
\end{equation}

Above, $u$ and $v$ are the horizontal velocities in the zonal and meridional directions, respectively. $w$ is the vertical velocity in the log-pressure coordinate system. $f=2\Omega\sin\vartheta$ is the Coriolis parameter. $\overline{D}$ symbolises the effects of friction and dissipation. $z^{\ast}=-H\log(p/p_0)$ is log-pressure `pseudoheight', with $H=RT_{\text{s}}/g$ a pressure scale-height and $p_0$ a reference pressure (e.g., the mean surface pressure). $\rho_0=\rho_{\text{s}}e^{-z^{\ast}/H}$ is a basic state density profile, where $\rho_{\text{s}}=p_0/(RT_{\text{s}})$. 
$\theta=T(p_0/p)^{R/c_p}$ is the potential temperature. $R$ is the specific gas constant, $c_p$ is the specific heat capacity at constant pressure, $T_{\text{s}}$ the globally-averaged surface temperature, and $g$ the acceleration due to gravity. A number of variables are decomposed into their zonal-average and a deviation from the zonal average, using the notation $X=\overline{X}+X^{\prime}$. 

We note that equation \eqref{eq:TEM_u} can be simplified considerably by re-writing it in terms of the axial component of angular momentum $m$, yielding \begin{equation}
    \rho_0\frac{\text{D}^{\ast}\overline{m}}{\text{D}t}=\nabla\cdot\boldsymbol{F}-\rho_0\overline{D}_{m}, \label{eq:TEM_m}
\end{equation}
with $D_{m}=a\cos\vartheta D$. From equation \eqref{eq:TEM_m} it is clear that if $D_m$ transports angular momentum down gradient, then up gradient transport can only be achieved by $\nabla\cdot\boldsymbol{F}$ (i.e., non-axisymmetric disturbances; cf. \citealp{1969JAtS...26..841H}). 

The TEM formulation of the zonally-averaged momentum budget has two main advantages over the regular Eulerian mean momentum budget \citep{1987mad..book.....A}. First, the residual mean circulation approximates that driven by diabatic heating, and thus more closely resembles the Lagrangian mean meridional mass flow. Residual mean meridional overturning can be quantified via definition of a streamfunction \begin{equation}
    \Psi^{\ast} = \frac{2\pi a\cos\vartheta}{g}\int_0^{p}\overline{v}^{\ast}\,\text{d}p.
\end{equation}
Second, in the quasigeostrophic limit, $\nabla\cdot\boldsymbol{F}$ describes the total wave-induced forcing of the TEM circulation. Additionally, $\nabla\cdot\boldsymbol{F}$ depends on the basic physical properties of wave or eddy disturbances in such a way that it is zero if the disturbances are steady, linear, frictionless and adiabatic.

\begin{figure*}[!ht]
    \centering\includegraphics[width=.85\textwidth]{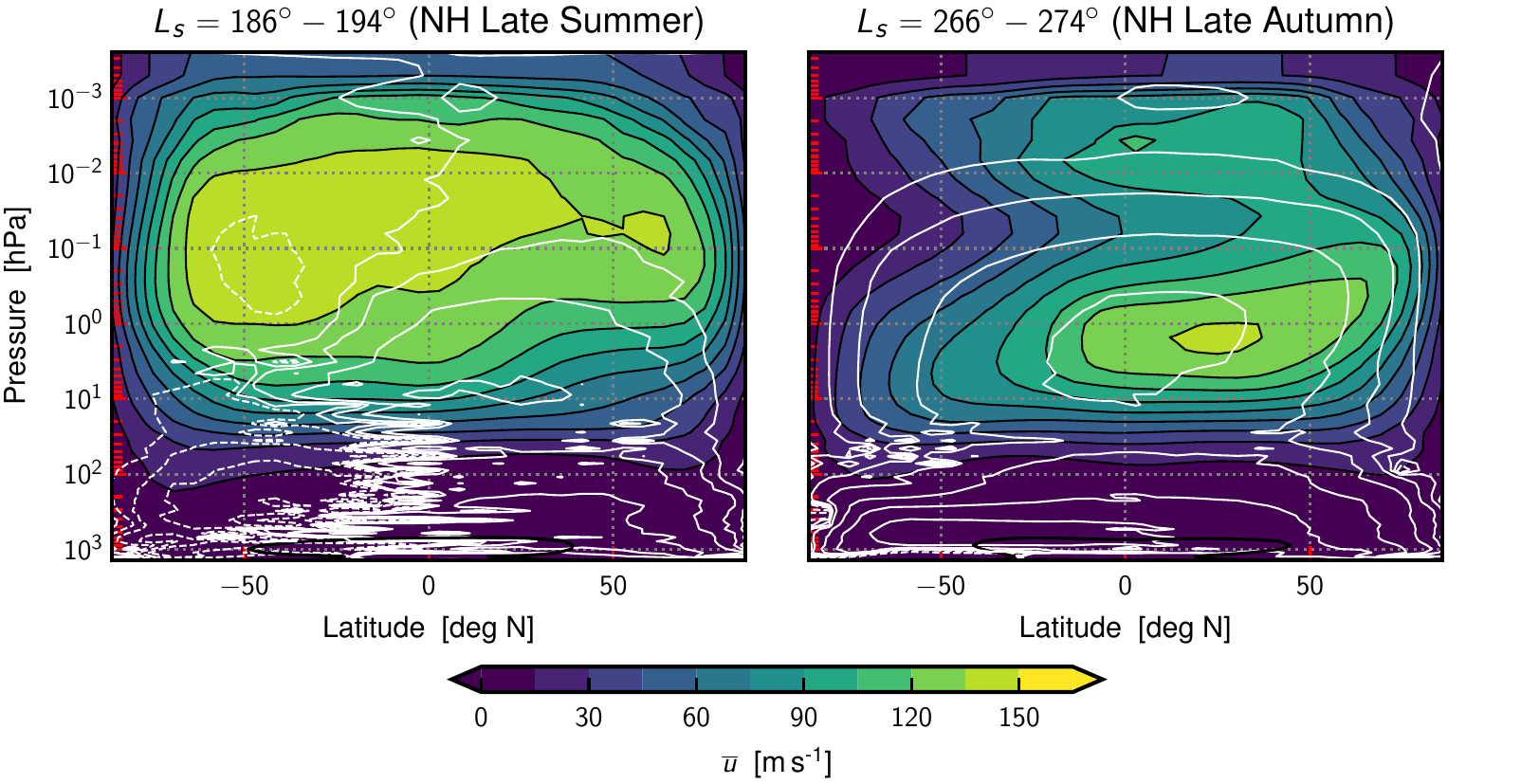}
    \caption{Zonal-mean zonal wind $\overline{u}$ and TEM streamfunction $\Psi^{\ast}$ for northern hemisphere late summer and late autumn. Six contour levels are shown for $\Psi^{\ast}$ at $\pm10^{7}$, $\pm10^{8}$ and $\pm10^{9}$ $\text{kg}\,\text{s}^{-1}$ (negative contours dashed). Positive $\Psi^{\ast}$ indicates clockwise overturning.}\label{fig:zm_fulldepth}
\end{figure*}

\subsubsection{Spectral diagnostics}

In order to characterise wave-like disturbances present in our model atmosphere, we use spectral analyses to transform time series data from real space ($t,\lambda$) to spectral space ($\omega,k$). Here, $t$ is time, $\lambda$ is longitude, $\omega$ is frequency and $k$ is the zonal wavenumber, related to the number of wavelengths $k_n$ that fit around a latitude circle via $k=k_n/(a\cos\vartheta)$. For some quantity $g(t,\lambda)$ defined in real space, this is achieved via the complex Fourier transform \begin{equation}
    \tilde{G}(\omega,k)=\frac{1}{2\pi T}\int^{T}_{0}\int^{2\pi}_{0}g(t,\lambda)e^{-i(k_n\lambda-\omega t)}\,\text{d}\lambda\,\text{d}t. 
\end{equation}
Here, $k$ (or $k_n$) is positive, and $\omega>0$ and $\omega<0$ correspond to eastward and westward propagating modes, respectively. Given time series $f$ and/or $g$, cross-spectral power density can then be computed via \begin{equation}
    C_{\omega,k} = 2\langle\,\text{Re}\lbrace\tilde{F}\tilde{G}^{\ast}\rbrace\,\rangle, \label{eq:cospec}
\end{equation}
where the asterisk denotes the complex conjugate, and angle brackets represent averaging over a frequency bandwidth. We follow the method of \citet{1971JMSJ...49..125H} for the computation of equation \eqref{eq:cospec}, and averaging in frequency is achieved by application of a Gaussian window following \citet{1991JAtS...48..688R}. { Spectra are computed from subsets of the data  $8^{\circ}$ $L_{\text{s}}$ in duration, corresponding to roughly $235\ \text{(Earth) days}$ (1705 data points). This yields a raw spectral resolution of $2\pi\times\Delta\omega = 4.69\times10^{-3}\ \text{days}^{-1}$, smoothed to an effective resolution of $2\pi\times \Delta\omega_{\text{eff}}=1.67\times10^{-2}\ \text{days}^{-1}$ by the Gaussian window.} 

It is worth noting that the absolute frequency, $\omega$, of a transient wave can be decomposed into a contribution from the intrinsic frequency of the wave, $\omega^{+}$, and a doppler shift due to the background zonal-mean zonal wind, so that \citep{1987mad..book.....A} \begin{equation}
    \omega^{+}=\omega-k\overline{u}.\label{eq:intrinsic}
\end{equation}
$\omega^{+}$ is then positive for waves that propagate eastward relative to the background zonal wind (prograde waves), and negative for waves that propagate westward relative to the background wind (retrograde waves). 

The propagation of waves relative to the mean flow can also be assessed by transforming spectra in frequency-wavenumber ($\omega,k$) space to phase speed-wavenumber ($c_p,k$) space, where $c_p=\omega/k$. This is achieved via the relation \begin{equation}
    C_{c_p,k}=kC_{\omega,k},
\end{equation}
again following \citet{1991JAtS...48..688R}. { The lower bounds on raw and effective resolution in phase-speed space are $\Delta c_{p}=0.879\,\text{m\,s}^{-1}$ and  $\Delta c_{p,\text{eff}}=3.13\,\text{m\,s}^{-1}$, respectively (corresponding to $\Delta\omega$ and $\Delta\omega_{\text{eff}}$ above for $k_{n}=1$ at the equator).} Phase speed spectra will be shown as a function of $c_p$ and some other spatial dimension $x$ (latitude or pressure). This is achieved by computing $C_{c_p,k}$ for each $x$ and then summing over wavenumber.

{ The composite spatial structure of selected waves is obtained using a common linear regression technique \citep{2000JAtS...57..613W,2013JCli...26..988A}. For a given disturbance of interest, a reference timeseries is constructed from the average value of a reference variable ($v^{\prime}$ for mixed Rossby--Gravity (MRG) waves \footnote{{ For all waves aside from MRG waves, the anomalous geopotential height, $z^{\prime}$, is used as the reference variable when constructing composites. For MRG waves, $z^{\prime}$ becomes small close to the equator, and so the meridional wind anomaly is used instead.}}, $z^{\prime}$ otherwise, where $z$ is the geopotential height) at the two latitude points located immediately north and south of the equator.  The reference variable at the equator is then filtered for frequencies and wavenumbers of interest by obtaining its Fourier transform, zeroing the resulting spectrum outside of the region of interest, and then transforming the variable back into real space. The resulting timeseries from all longitude points are then concatenated to create the reference timeseries (this is justified as the model's lower boundary condition is zonally symmetric).
To obtain the composite spatial structure, unfiltered fields are then regressed against the reference timeseries (after a similar concatenation in longitude). This produces a spatial field of regression coefficients, which are then rescaled by twice the standard deviation of the reference timeseries to create the final composite. In this work, we only produce composites for disturbances with zonal wavenumber $k_{n}=1$, and frequencies are filtered to retain those falling within the range $\Delta\omega = \Delta c_{p}k$, where $\Delta c_{p}$ is centred on the phase speed of interest. $\Delta c_{p}$ is chosen to have a width of $10\,\text{m\,s}^{-1}$, which corresponds to roughly 11 frequency bins in Fourier space. }

\section{Basic circulation and seasonal cycle}\label{sec:basic_circulation}

Before proceeding with our analysis of the waves present in our simulation of Titan's atmosphere, it is useful to discuss some aspects of the simulation's zonally-averaged circulation (on which the waves themselves depend). 

Figure \ref{fig:zm_fulldepth} shows the circulation averaged over two periods close to the northern hemisphere late summer ($L_{\text{s}} = 180^{\circ}$ -- $190^{\circ}$) and late autumn ($L_{\text{s}} = 260^{\circ}$ -- $280^{\circ}$), respectively (representative of equinoctal and solstitial conditions). Colour contours show the zonal-mean zonal wind, $\overline{u}$, and white contours show the TEM streamfunction, $\Psi^{\ast}$. The circulation displays a clear dependence on the seasonal cycle, most pronounced in the stratosphere (roughly $p>40\,\text{hPa}$). Around equinox, the circulation has some hemispheric symmetry, whereas around the solstice it is very asymmetric, with $\overline{u}$ strongest in the winter hemisphere, and $\Psi^{\ast}$ taking the form of a single cell rising near the summer pole and sinking near the winter pole. The structure of $\Psi^{\ast}$ is similar to that of the regular Eulerian mean streamfunction computed in previous studies \citep[e.g.,][]{2019Icar..333..113L,2023Icar..39015291L}; this is due to the meridional eddy heat flux being small (so $\psi$ in Section \ref{sec:methods} is small, and $\overline{v}^{\ast}\approx\overline{v}$). Also noteworthy is a pronounced shear in the zonal wind, in both the vertical and meridional directions. The vertical shear is strongest in the range $p=10$ -- $100\,\text{hPa}$ and does not vary that substantially with the seasonal cycle. Meridional shear is most pronounced around $p=1\,\text{hPa}$ and is strongest around the solstice.

\begin{figure}
    \centering\includegraphics[width=\linewidth]{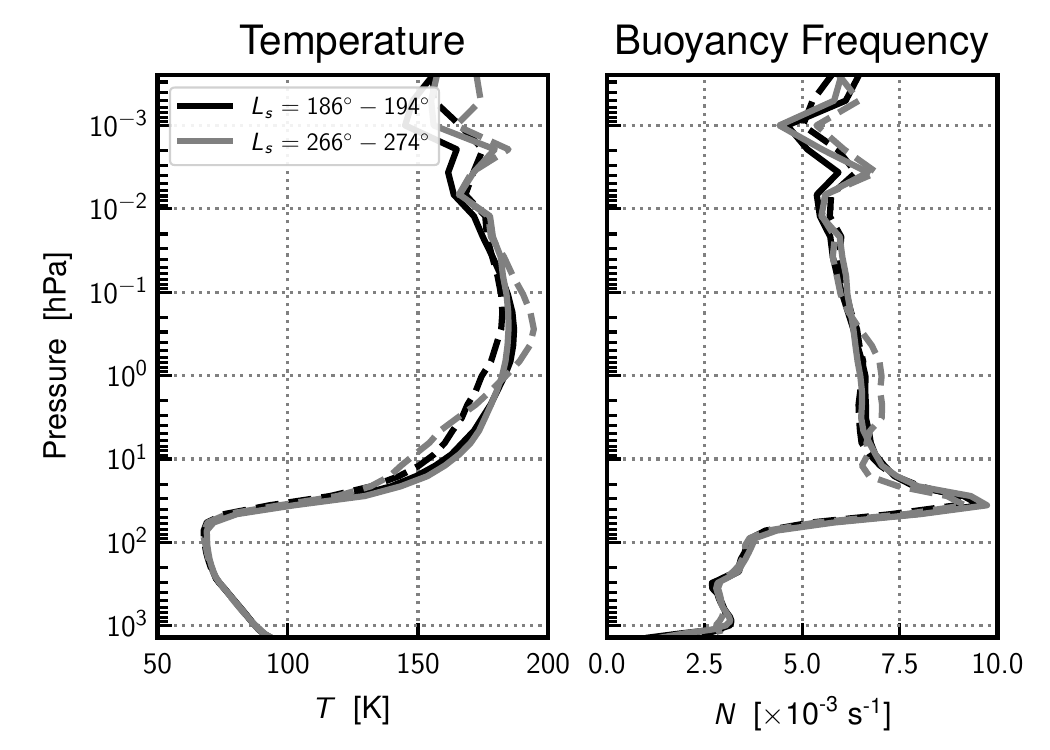}
    \caption{Vertical profile of temperature $T$ and buoyancy frequency $N$ at the equator (solid) and $60^{\circ}\,\text{N}$ (dashed), around northern hemisphere late summer (black) and late autumn (grey).}\label{fig:NT}
\end{figure}

\begin{figure*}[!t]
    \centering\includegraphics[width=.85\textwidth]{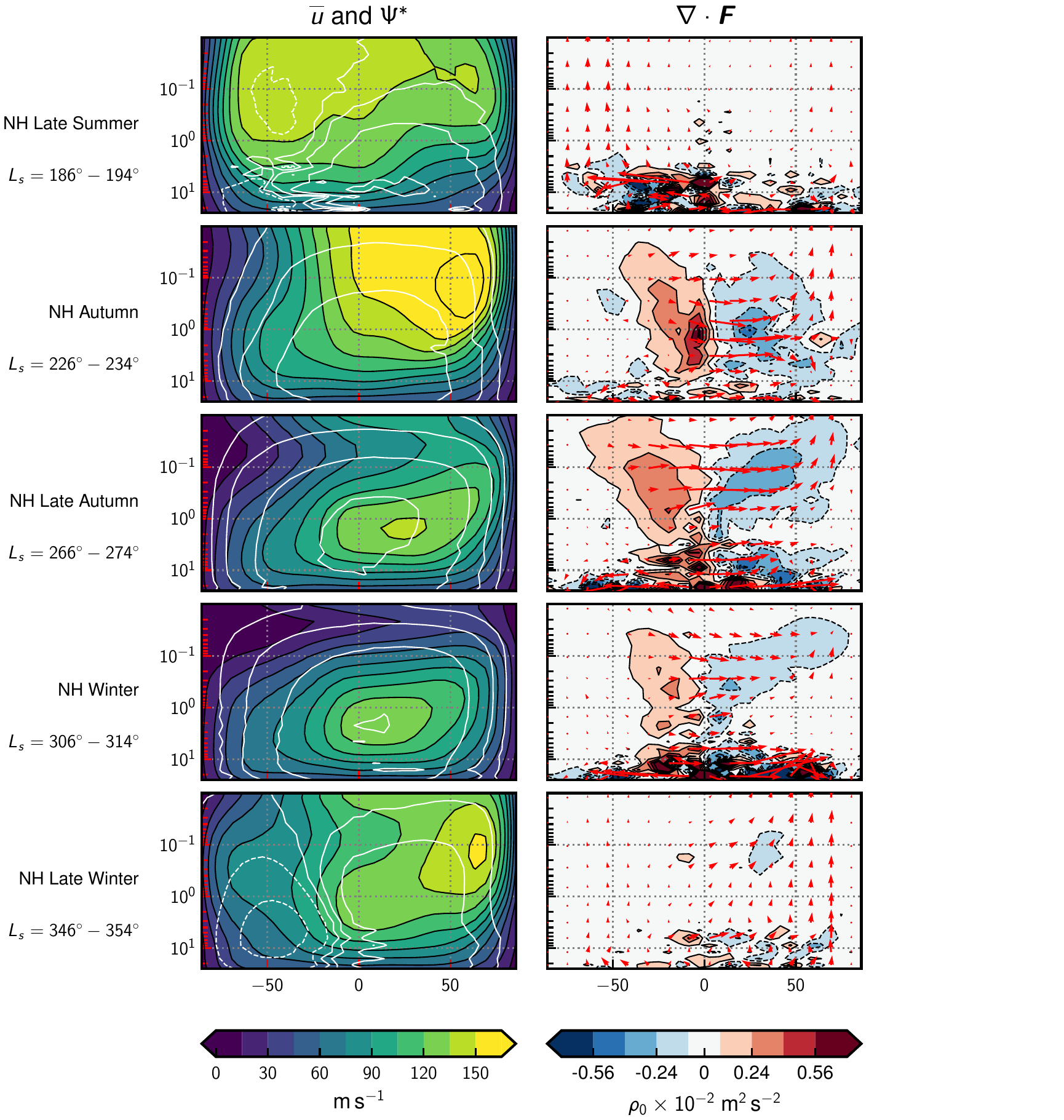}
    \caption{Seasonal cycle of $\overline{u}$, $\Psi^{\ast}$, $\boldsymbol{F}$ and $\nabla\cdot\boldsymbol{F}$ across the pressure range $p=0.04$ -- $40\,\text{hPa}$. Left column: Zonal-mean zonal wind $\overline{u}$ and TEM streamfunction $\Psi^{\ast}$, as in Figure \ref{fig:zm_fulldepth}. Right column: the EP flux $\boldsymbol{F}$ (quivers) and its divergence $\nabla\cdot\boldsymbol{F}$ (colour contour). The horizontal and vertical components of $\boldsymbol{F}$ are rescaled as $(\pi a)^{-1}F^{(\vartheta)}$ and $(\Delta z^{\ast})^{-1}F^{(z^{\ast})}$, respectively, where $\Delta z^{\ast}=-H\log(p_{\text{t}}/p_{\text{b}})$ with $p_{\text{t}}=2\times10^{-2}\,\text{hPa}$ and $p_{\text{b}}=20\,\text{hPa}$. The EP flux divergence has units of density $\times$ specific angular momentum per unit time. Red colour contours indicate acceleration.}\label{fig:seasonal_TEM}
\end{figure*}

The temperature structure of an atmosphere also has important effects on the characteristics of atmospheric waves, for example through the dependence of the deformation radius (see Appendix \ref{sec:appendix}) on the buoyancy frequency $N$, defined as \citep{1987mad..book.....A} \begin{equation}
    N=\left[\frac{R}{H}\left(\frac{\partial T}{\partial z^{\ast}}+\frac{\kappa T}{H}\right)\right]^{\nicefrac{1}{2}}\kern-10pt, 
\end{equation}
where $\kappa=R/c_{\text{p}}$. Figure \ref{fig:NT} shows $T$ and $N$ vs. $p$, at the equator and in mid-latitudes, for northern hemisphere autumn equinox and winter solstice. {A `cold point' tropopause is located at roughly $p=60\,\text{hPa}$,} and below $60\,\text{hPa}$, temperature decreases with height in a region that could be termed the troposphere. In the troposphere, there is little seasonal variation in $T$ (and so likewise $N$), and additionally meridional temperature contrasts are very weak \citep{2016AREPS..44..353M}. In the stratosphere, temperature increases sharply with height until $p\approx1\,\text{hPa}$, above which the atmosphere is roughly isothermal. Meridional temperature contrasts are somewhat more pronounced  in the stratosphere when compared with the troposphere, and there is also some evidence for seasonal variation of the temperature structure, as in observations \citep{2008Icar..194..263A,2012Icar..221.1020S}. The sharp increase in temperature in the lower stratosphere notably gives rise to a pronounced increase in $N$ between $p=60\,\text{hPa}$ and $p=20\,\text{hPa}$.

\begin{figure*}
    \centering\includegraphics[width=.85\textwidth]{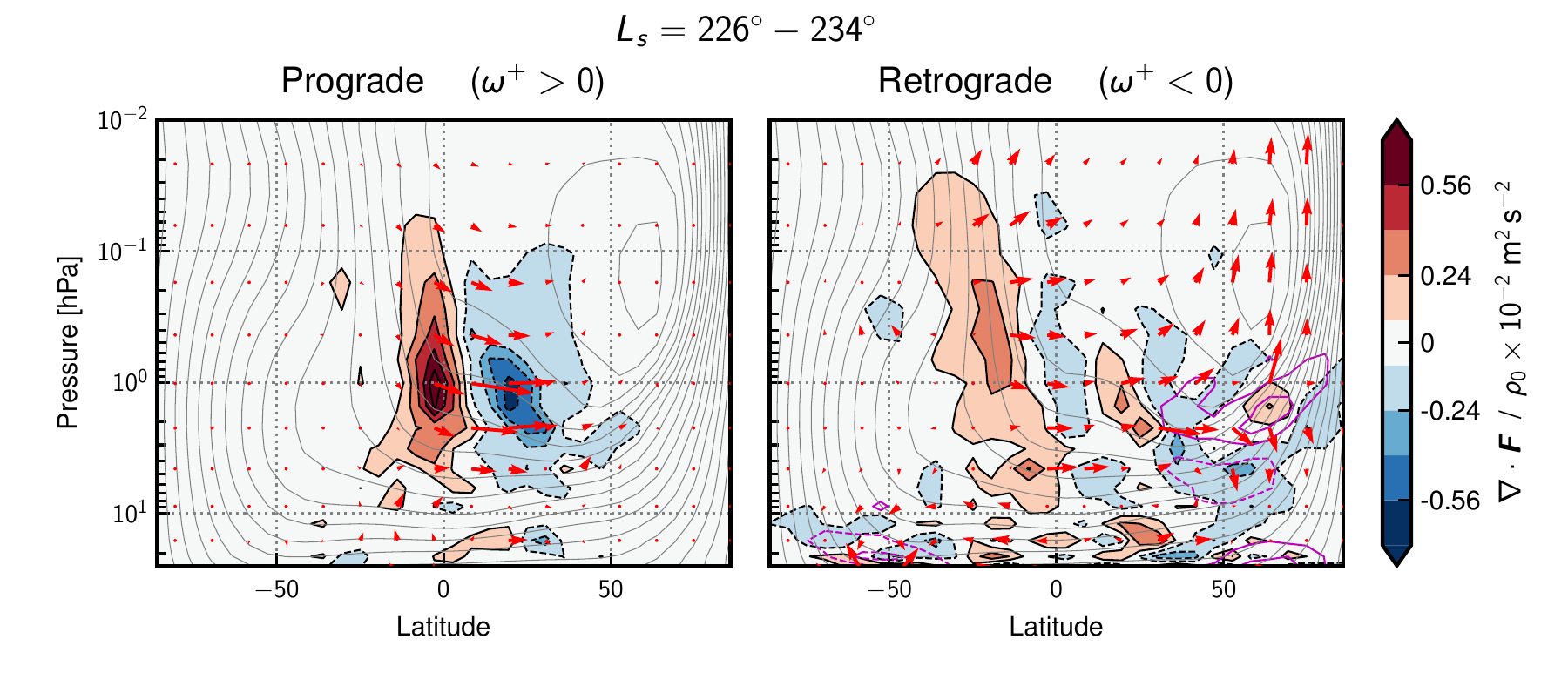}
    \caption{EP flux $\boldsymbol{F}$ (quivers) and its divergence $\nabla\cdot\boldsymbol{F}$ (colour contour) for prograde (left) and retrograde (right) waves during the period $L_{\text{s}} = 226^{\circ}$ -- $234^{\circ}$. $\boldsymbol{F}$ is scaled as in Figure \ref{fig:seasonal_TEM}. Grey contours show the zonal-mean zonal wind $\overline{u}$ with the same contour levels as in Figures \ref{fig:zm_fulldepth} and \ref{fig:seasonal_TEM}. Purple contours show the contribution to $\nabla\cdot\boldsymbol{F}$ from $\partial F^{(z^{\ast})}/\partial z^{\ast}$ (using the same contour levels).}\label{fig:EP_eastwest1}
\end{figure*}

Figure \ref{fig:seasonal_TEM} presents a more granular view of seasonal variation in the zonally-averaged circulation { over the pressure range $p=0.025$ -- $25\,\text{hPa}$.} The left-hand column shows $\overline{u}$ and $\Psi^{\ast}$, as in Figure \ref{fig:zm_fulldepth}, and the right-hand column shows the EP flux (quivers) and its divergence $\nabla\cdot\boldsymbol{F}$ (colours). 

\autoref{fig:seasonal_TEM} shows that through much of the year, the stratospheric zonal-mean circulation is strongly asymmetric about the equator. Over the period  $L_{\text{s}} = 226^{\circ}$ -- $314^{\circ}$ the TEM streamfunction indicates that overturning takes the form of a single cell rising in the summer hemisphere and sinking in the winter hemisphere, and only for a brief period around equinox is a two cell structure apparent. Likewise, throughout most of the year zonal wind velocities are clearly much stronger in the winter hemisphere{, with the greatest hemispheric asymmetry occurring in mid-autumn (around $L_{\text{s}}=230^{\circ}$). }

{ Across the pressure range $p=0.1$\,--\,$10\,\text{hPa}$, the magnitude and meridional structure of the zonal-mean zonal wind, and its seasonal variation, is comparable with that inferred from Cassini/CIRS temperature measurements (via gradient wind balance; \citealp{2021Icar..35414030S}, their Figure 7), for which a maximum velocity at $p=0.1\,\text{hPa}$ of roughly $180\,\text{m\,s}^{-1}$ is obtained in southern hemisphere autumn ($L_{\text{s}}=50^{\circ}$; their observations do not cover northern hemisphere autumn). The agreement between model and Cassini/CIRS-derived wind speeds is less good above $p=0.1\,\text{hPa}$, where the observed zonal velocities (in excess of $200\,\text{m\,s}^{-1}$) are generally greater than those obtained in the simulation. This is due to the effect of the model sponge, which acts to decelerate the winds near the model top. A detailed comparison between TAM's simulated circulation and observations is presented in \citet{2023Icar..39015291L}. }

Throughout much of the year, and across a range of pressures, the EP flux divergence is positive at or near the equator, consistent with the requirement for superrotation to be maintained by eddy accelerations. Seasonal variation is also evident in the EP flux and its divergence. This is particularly true in the upper stratosphere ($p\lesssim2\,\text{hPa}$), where, during northern hemisphere mid-autumn to mid-winter, eddies induce a southward flux of angular momentum that balances northward momentum transport by the mean flow, whereas around the equinoxes eddy acceleration is greatly reduced. Lower down in the stratosphere ($p\gtrsim2\,\text{hPa}$), eddy accelerations display weaker seasonal variation.

The direction of the EP flux vectors is indicative of the direction of wave propagation, although whether $\boldsymbol{F}$ is roughly parallel or anti-parallel to the wave propagation depends on whether a wave has a prograde or retrograde intrinsic frequency \citep{1987mad..book.....A,2006JAtS...63.1623I}. Nonetheless, it is clear from Figure \ref{fig:seasonal_TEM} that waves propagate both vertically and horizontally. We note that for waves that have a horizontal size close to the planetary scale, `horizontal propagation' should be taken to mean that the waves are preferentially generated at one latitude and dissipated at another (without necessarily physically propagating in the meridional direction). In regions where the EP flux divergence is strongest, $\boldsymbol{F}$ is very close to horizontal, and a decomposition of $\nabla\cdot\boldsymbol{F}$ into contributions from $F^{(\vartheta)}$ and $F^{(z^{\ast})}$ reveals that eddy acceleration is almost entirely due to horizontal divergence (not shown).

{In the next section, we characterise the waves present in our model atmosphere, and identify those that contribute most to the acceleration of superrotation. We focus exclusively on northern hemisphere autumn, during which there is significant eddy momentum flux divergence in both the upper and lower stratosphere. In the subsequent discussion, we briefly describe the similarities and differences between the disturbances we identify for autumn, and those present during the rest of the year.}

\section{Equatorial waves and acceleration of superrotation}\label{sec:autumn}

\begin{figure*} 
    \centering\includegraphics[width=.9\textwidth]{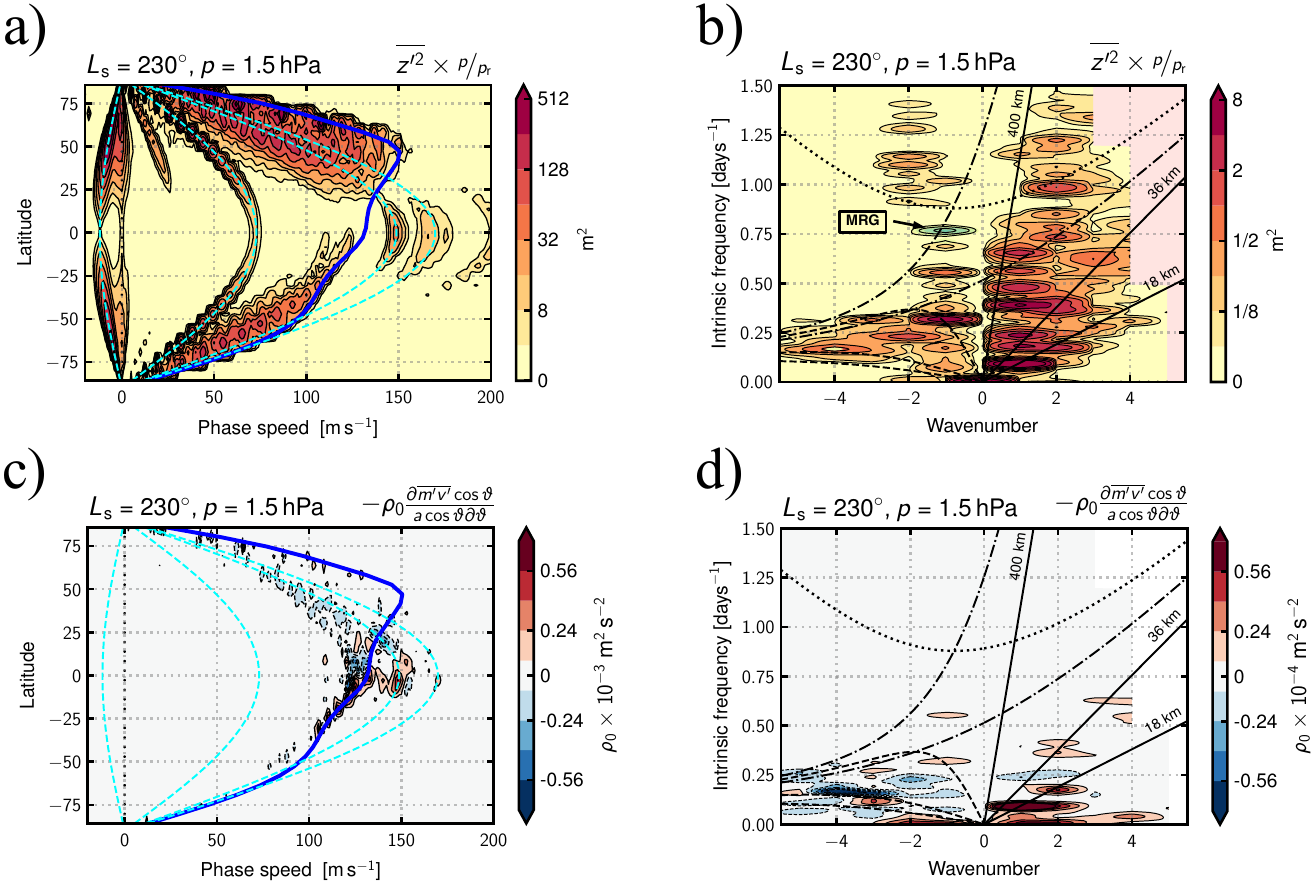}
    \caption{{ Spectra of} eddy geopotential height and horizontal angular momentum flux convergence presented as a function of latitude and phase speed (panels a and c), and wavenumber and frequency (panels b and d), averaged between $\pm6^{\circ}$ latitude. Analysis is shown for $p=1.5\,\text{hPa}${, $L_{\text{s}} = 226^{\circ}$ -- $234^{\circ}$.} The dashed cyan lines in panels a) and c) are curves of constant angular velocity ($c_p\times\cos\vartheta$), and solid blue lines show the zonal-mean zonal wind $\overline{u}$. In panels b) and d), prograde modes are plotted with a positive wavenumber and retrograde modes are plotted with a negative wavenumber. Blue colour in panel b) indicates modes anti-symmetric across the equator (only one is present, a mixed Rossby--gravity mode at $\omega^{+}=0.76\,\text{days}$ and $k_{n}=1$; labeled in panel). The curves in panels b) and d) are dispersion curves for Kelvin waves (solid), mixed Rossby--gravity / $n=0$ eastward inertio--gravity waves (dash-dot), $n=1$ inertio--gravity waves (dotted), and $n=1$ equatorial Rossby waves (dashed), where $n$ is the meridional mode number. { Curves are shown for vertical wavelengths $\lambda_{\text{v}}=400\,\text{km}$ and $36\,\text{km}$ for all waves, and additionally $\lambda_{\text{v}}=18\,\text{km}$ for equatorial Rossby and Kelvin waves.} These curves intersect with the $c_{p}=170\,\text{m\,s}^{-1}$ Kelvin wave, $c_{p}=149\,\text{m\,s}^{-1}$ Rossby--Kelvin wave, and $c_{p}=73\,\text{m\,s}^{-1}$ equatorial Rossby wave, shown in Figure \ref{fig:waves1}, respectively (all $k_n=1$). The dispersion curves are calculated assuming a rotation rate $\Omega_{\text{eff}}=\overline{u}/a$, which is the effective rotation rate moving with the superrotating jet. For details of the computation, see Appendix \ref{sec:appendix}.}\label{fig:spec1}
\end{figure*}

In this section, we analyse the waves present during northern hemisphere autumn, during which eddy acceleration of the zonal-mean flow is strongest.  { All of our analysis will focus on the  period  $L_{\text{s}} = 226^{\circ}$ -- $234^{\circ}$}. Analysing a short period makes for simpler spectral analysis, as it reduces the effect of a time--varying doppler shift of the waves due to seasonal variation in $\overline{u}$. We have chosen to discuss the period $L_{\text{s}} = 226^{\circ}$ -- $234^{\circ}$ because it yielded the cleanest results as far as visual presentation is concerned. However, we stress that the results described in this section are characteristic of the entire period  $L_{\text{s}} = 226^{\circ}$ -- $314^{\circ}$.

We begin by returning to the EP flux and its divergence, which we now separate into contributions from prograde and retrograde disturbances. This is shown in Figure \ref{fig:EP_eastwest1}. Both prograde and retrograde disturbances contribute to $\nabla\cdot\boldsymbol{F}$, but in different locations. Prograde waves mostly accelerate the mean flow in the upper stratosphere, whereas in the lower stratosphere the contribution from retrograde waves is enhanced. The purple contours indicate the contribution of $\partial F^{(z^{\ast})}/\partial z^{\ast}$ to $\nabla\cdot\boldsymbol{F}$, which is generally weak, and acceleration of the mean flow by eddies is identified to be mostly due to horizontal convergence of eddy angular momentum fluxes.

The direction of $\boldsymbol{F}$ informs us of the predominant direction of wave propagation for each case. Noting that $\boldsymbol{F}$ is aligned with the direction of propagation for retrograde waves, and opposite to the direction of propagation for prograde waves \citep{1987mad..book.....A,2006JAtS...63.1623I}, we identify that prograde waves appear to propagate equatorwards and, to a lesser extent, upwards from source regions in the middle stratosphere around $p=2\,\text{hPa}$ in northern hemisphere sub-tropical latitudes. Meanwhile, retrograde waves propagate polewards away from source regions close to the equator, and then vertically {(both upwards and downwards)} at mid- and high-latitudes.

To learn more about the specific wave modes that contribute to the general behaviour shown in Figure \ref{fig:EP_eastwest1}, we now perform a spectral decomposition of the perturbation fields $u^{\prime}$, $v^{\prime}$ and $z^{\prime}$ using the techniques described in Section \ref{sec:methods}. 

\subsection{Upper stratosphere}\label{sec:upper}

We begin by analysing the waves present in the upper stratosphere, focusing on { the pressure level $p=1.5\,\text{hPa}$.} Figure \ref{fig:spec1} shows latitude--phase speed, and frequency--wavenumber (averaged between $\pm6^{\circ}$ latitude), spectra for the eddy geopotential height anomaly and eddy angular momentum flux convergence. 

From Figure \ref{fig:spec1}a, we identify that both prograde ($c_p>\overline{u}$) and retrograde ($c_{p}<\overline{u}$) disturbances are present at the equator. A number of structures have coherence along curves of constant angular velocity. This is particularly clear for a series of retrograde modes with broad meridional extent ({ two examples of which are indicated by dashed cyan curves of constant angular velocity, with $c_p=-12\,\text{m\,s}^{-1}$ and $c_p=73\,\text{m\,s}^{-1}$ at the equator}). Figure \ref{fig:spec1}b shows that power in the equatorial disturbances is generally concentrated into low wavenumbers, but across a broad range of frequencies (although we note that the large mean flow contribution to $\omega$ means that disturbances with moderate eastward intrinsic frequencies $\omega^{+}$ and wavenumbers, i.e. the upper-right hand corner of Figure \ref{fig:spec1}b and d, are not resolved in our output data).

\begin{figure*}
    \centering\includegraphics[width=.85\textwidth]{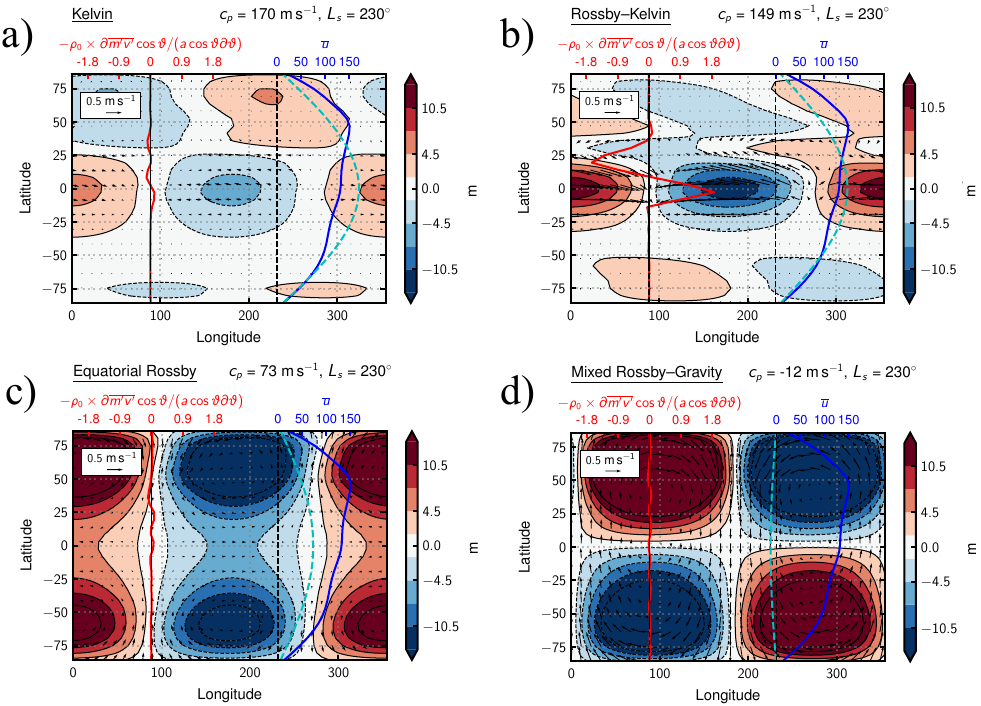}
    \caption{{ Composites of the} horizontal structure for various waves identified at $p=1.5\,\text{hPa}$ during the period { $L_{\text{s}} = 226^{\circ}$ -- $234^{\circ}$}. Eddy velocity components ($u^{\prime}$,$v^{\prime}$) are shown as quivers, and eddy geopotential height is shown as colour contours. The quiver scale is constant between panels. Each panel shows fields filtered for $k_n=1$ and frequencies corresponding to $c_p=c_{p,\text{wave}}\pm c_{p,\text{diff}}$. $c_{p,\text{wave}}$ for each wave is indicated above each panel, and corresponds to one of the cyan lines in Figure \ref{fig:spec1}a,c;  $c_{p,\text{diff}}=5\,\text{m\,s}^{-1}$.  The red line in each panel shows the horizontal angular momentum flux convergence computed directly from each wave's filtered eddy velocity field. It has units of $\rho_0\times10^{-3}\,\text{m}^{2}\,\text{s}^{-2}$. Blue lines show the zonal-mean zonal wind $\overline{u}$, and the dashed cyan line shows $c_p\times\cos\vartheta$ (both with units $\text{m}\,\text{s}^{-1}$).}\label{fig:waves1}
\end{figure*}

In Figure \ref{fig:spec1}b, retrograde waves are plotted with a negative wavenumber, while prograde waves are plotted with a positive wavenumber. The $\omega^{+}$--$k$ distribution of $\overline{z^{\prime2}}$ suggests that a number of equatorial wave modes are supported in the model atmosphere, including low frequency equatorial Rossby waves (retrograde) and Kelvin waves (prograde), as well as higher frequency inertio-gravity modes (retrograde and prograde). { No background spectrum has been removed to minimise the effects of random or nonperiodic processes on the spectrum shown in Figure \ref{fig:spec1}b  \citep{1999JAtS...56..374W}, which implies that synoptic scale waves dominate over mesoscale disturbances in the model atmosphere. Moreover, the spectra are highly discretized in wavenumber and frequency, which indicates that the modes are circumglobal, as opposed to being organised into localised wave packets (see, e.g., \citealp{2021JAMES..1302528T}).} Colour in Figure \ref{fig:spec1}b is used to differentiate between modes that are symmetric (red) and anti-symmetric (blue) across the equator (see \citealp{1999JAtS...56..374W} for method). Almost all disturbances are symmetric, with the only notable exception being an anti-symmetric mode with $\omega^{+}=-0.76\,\text{days}^{-1}$ and $k_n=1$. This corresponds to the structure at $c_p=-12\,\text{m\,s}^{-1}$ shown in Figure \ref{fig:spec1}a, which we will later identify as a mixed Rossby--gravity wave (Figure \ref{fig:waves1}d). 

To determine which of these modes are involved in inducing equatorward angular momentum transport, we now turn our attention to Figures \ref{fig:spec1}c and d, which present a spectral decomposition of the eddy angular momentum flux divergence. The high-frequency inertio-gravity waves and the mixed Rossby-gravity wave do not appear to play a role in the acceleration of superrotation. Instead, the largest eddy accelerations are due to modes with phase speeds similar to the mean flow speed (i.e., small $\omega^{+}$) and very low wavenumbers $k_n=1,2$. Generally these disturbances have $c_p>\overline{u}$ at latitudes where $\overline{m^{\prime}v^{\prime}}$ is convergent,  and $c_p<\overline{u}$ where $\overline{m^{\prime}v^{\prime}}$ is divergent.

Figure \ref{fig:waves1} shows the horizontal structures of selected waves identified in Figure \ref{fig:spec1}a,c by dashed cyan lines. Two prograde modes and two retrograde modes are presented. In each panel, composites of $u^{\prime}$ and $v^{\prime}$ are shown as quivers, and composites of $z^{\prime}$ as colour contours. The composites are obtained following the method described in Section \ref{sec:methods}. { For simplicity, we restrict our analysis to modes with $k_n=1$. Then, reference timeseries for the composites are obtained by filtering either $z^{\prime}$ or $v^{\prime}$ (for MRG waves) for frequencies in the range $\Delta\omega = \Delta c_{p} / a = (c_{p}\pm c_{p,\text{diff}})/a$, where $c_{p}$ is the disturbance's phase speed at the equator, and $c_{p,\text{diff}}=5\,\text{m\,s}^{-1}$.}

Figure \ref{fig:waves1}a and b present filtered fields for the phase speeds $c_p=170\,\text{m\,s}^{-1}$ and $c_p=149\,\text{m\,s}^{-1}$. The waves in these panels have a similar structure, composed of a prograde equatorial Kelvin wave-like component on the equator, phase-locked with a retrograde, Rossby wave-like component in northern mid-latitudes \citep[cf.][]{1966JMeSJ..44...25M,2009RvGeo..47.2003K}. This sort of wave mode has been identified in previous studies of superrotating atmospheres, and is often referred to as a `Rossby--Kelvin' mode \citep{2005JAtS...62.2514I,2010JGRE..11512008M,2014GeoRL..41.4118W}. The Rossby--Kelvin waves are flanked by extratropical Rossby waves (north of ${\sim}\,50^{\circ}$ for $c_p=170\,\text{m\,s}^{-1}$ and ${\sim}\,40^{\circ}$ for the $c_p=149\,\text{m\,s}^{-1}$ wave). The $c_p=170\,\text{m\,s}^{-1}$ Rossby--Kelvin wave has a larger meridional extent ($\pm50^{\circ}$) than the $c_p=149\,\text{m\,s}^{-1}$ wave ($\pm35^{\circ}$) which is consistent with the prediction of linear theory when the intrinsic frequency $\omega^{+}$ is reduced \citep{1987mad..book.....A}. These modes correspond to the Kelvin modes in Figure \ref{fig:spec1}b with wavenumber $1$, and frequencies $\omega^{+}=0.21\,\text{days}^{-1}$ and $\omega^{+}=0.096\,\text{days}^{-1}$, respectively, { which reside on the $\lambda_{\text{v}}=36\,\text{km}$ and $\lambda_{\text{v}}=18\,\text{km}$ dispersion curves.} The equatorial deformation radii $l_{\text{d}}$ predicted by linear theory for these modes are $l_{\text{d}}/a=0.74$, corresponding to $\pm43^{\circ}$ latitude, for the $c_p=170\,\text{m\,s}^{-1}$ mode, and $l_{\text{d}}/a=0.52$, corresponding to $\pm31^{\circ}$ latitude, for the $c_p=149\,\text{m\,s}^{-1}$ mode, which agrees relatively well with the meridional extent of each mode inferred from Figure \ref{fig:waves1} (see Appendix \ref{sec:appendix} for details of the computation for $l_{\text{d}}$). 

\begin{figure*}
    \centering\includegraphics[width=.9\textwidth]{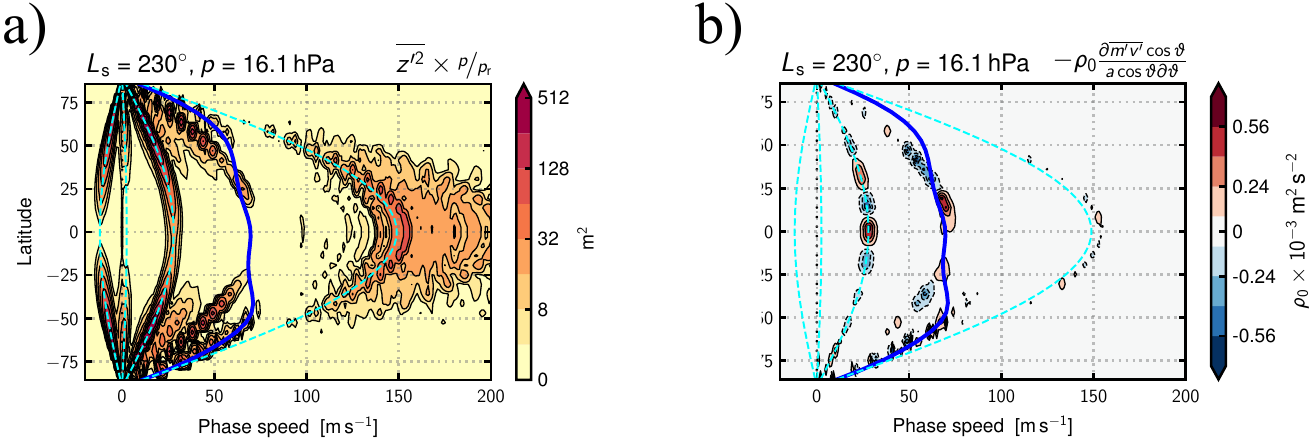}
    \caption{{ Spectra of} eddy geopotential height and horizontal angular momentum flux convergence presented as a function of latitude and phase speed, as in Figure \ref{fig:spec1}a,c but for $p=16.1\,\text{hPa}$. The dashed cyan lines are curves of constant angular velocity ($c_p\times\cos\vartheta$), and solid blue lines show the zonal-mean zonal wind $\overline{u}$.} \label{fig:spec2}
\end{figure*}

Phase locking between the prograde and retrograde components is made possible by the meridional shear in the background wind (see the solid blue line in each panel), and due to the asymmetry of the background shear, the combined Rossby--Kelvin mode is asymmetric about the equator, with no Rossby-like component in the southern hemisphere. Spatial overlap between the Rossby wave-like and Kelvin wave-like components is sufficiently large for the $c_p=149\,\text{m\,s}^{-1}$ wave to generate momentum flux convergence at the equator, whereas for the $c_p=170\,\text{m\,s}^{-1}$ wave the overlap is small and hence the equatorial momentum flux convergence is negligible \citep{2014GeoRL..41.4118W}. We note that we have identified additional, faster Kelvin waves with $c_p\gtrsim200\,\text{m\,s}^{-1}$ that do not have a Rossby-like component in middle latitudes (e.g., the prograde disturbance with equatorial phase speed $c_p\approx200\,\text{m\,s}^{-1}$ in Figure \ref{fig:spec1}a; spatial structure not shown), likely because their high phase speed prevents phase locking with a retrograde component in sub-tropical or middle latitudes.

Two retrograde modes are presented in Figure \ref{fig:waves1}.  One (Figure \ref{fig:waves1}c; $c_p=73\,\text{m\,s}^{-1}$) has a horizontal structure that is characteristic of an equatorial Rossby wave \citep{1966JMeSJ..44...25M,2009RvGeo..47.2003K}. The wave corresponds to the wavenumber 1 feature in Figure \ref{fig:spec1}b with intrinsic frequency $\omega^{+}=-0.31\,\text{days}^{-1}$. { It has a vertical wavelength of} $\lambda_{v}\approx400\,\text{km}$, 
associated with a theoretical deformation radius corresponding to a meridional extent of $\pm101^{\circ}$ latitude (i.e., the simulated mode has grown to fill the domain). The second mode shown in Figure \ref{fig:waves1}d has the structure of a mixed Rossby--gravity wave \citep{2009RvGeo..47.2003K}, which travels with a much higher intrinsic frequency than the equatorial Rossby wave. The effect of each of these retrograde waves on the angular momentum budget in the equatorial area is minimal compared with that of the Rossby--Kelvin wave shown in Figure \ref{fig:waves1}b, which is consistent with the separation of $\nabla\cdot\boldsymbol{F}$ shown in Figure \ref{fig:EP_eastwest1}, where prograde modes are responsible for the majority of eddy momentum flux convergence near the equator in the upper stratosphere. 

\subsection{Lower stratosphere}

\begin{figure*}
    \centering\includegraphics[width=.85\textwidth]{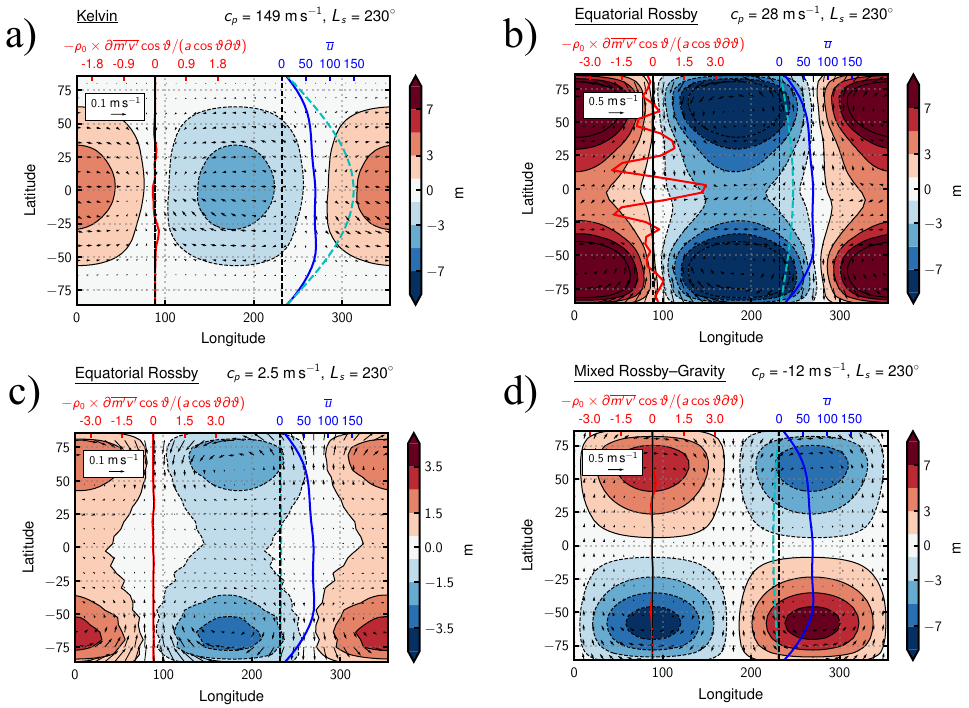}
    \caption{{ Composites showing the} horizontal structure for various waves as in Figure \ref{fig:waves1} but for $p=16.1\,\text{hPa}$. For panels b)--d) the quiver scale is constant, and the same as in Figure \ref{fig:waves1}. For panels a) and c) the quiver length has been multiplied a factor of 5. As before, the filtered frequencies correspond to $c_p=c_{p,\text{wave}}\pm c_{p,\text{diff}}$. $c_{p,\text{wave}}$ is indicated above each panel, and $c_{p,\text{diff}}=5\,\text{m\,s}^{-1}$.}\label{fig:waves2}
\end{figure*}

We now turn our attention to waves in the lower stratosphere, { focusing on the pressure level $p=16.1\,\text{hPa}$.} Figure \ref{fig:spec2} shows latitude--phase speed spectra of $\overline{z^{\prime2}}$ and the horizontal eddy angular momentum flux convergence, as in Figure \ref{fig:spec1}a,c but for the lower stratosphere. The picture is quite different to that in the upper stratosphere. First, zonal-mean zonal wind is more symmetric about the equator, and this appears to be reflected in the spectra shown in Figure \ref{fig:spec2}. Prograde and retrograde disturbances are both present (Figure \ref{fig:spec2}a), as before, but eddy angular momentum flux convergence is now primarily due to retrograde waves. It is notable that prograde modes appear to exist at phase speeds similar to those in the upper atmosphere, implying that the Kelvin waves may have a somewhat coherent vertical structure. A similar observation can also be made at the phase speed of the mixed Rossby--gravity wave identified previously ($c_p=-12\,\text{m\,s}^{-1}$). The vertical structure of the waves will be analysed in detail in the next subsection.

The horizontal structures of selected disturbances (indicated by cyan lines in Figure \ref{fig:spec2}) are shown in Figure \ref{fig:waves2}. The prograde disturbance with $c_p=149\,\text{m\,s}^{-1}$ has a structure that is very reminiscent of the solution for a linear Kelvin wave on the equatorial beta plane \citep{1966JMeSJ..44...25M,2009RvGeo..47.2003K}, aside from the fact that it has non-zero meridional velocity; this is likely a product of the spherical geometry \citep{2019Icar..322..103Y}. Notably, this mode is not accompanied by a Rossby wave-like component in mid-latitudes, unlike the disturbances with similar $c_p$  identified at $p=1.5\,\text{hPa}$. This underscores the importance of horizontal shear flow in the upper stratosphere for generating these features; without meridional shear that is comparable in magnitude to the phase speed of the prograde wave, prograde and retrograde components are unable to phase-lock \citep{2014GeoRL..41.4118W}. In the absence of an accompanying Rossby-like component, this mode is unable to converge momentum towards the equator as might be expected for predominantly divergent motion \citep{2005JAtS...62.2514I}.

Figure \ref{fig:waves2}b shows the retrograde wave which is responsible for horizontal eddy angular momentum convergence at the equator. This wave has a structure characteristic of an equatorial Rossby wave. The wave exhibits a slight equatorward meridional phase tilt close to the equator (see, e.g., the direction of the velocity quivers nearest to the equator), possibly induced by vertical shear in the background zonal-mean wind \citep{2006JAtS...63.1623I}. This subtle phase tilt allows the wave to converge momentum towards the equator. \citet{2016JAtS...73.3181D} have shown that similar disturbances are responsible for accelerating equatorial superrotation in idealised GCM simulations of an Earth-like planet with a smaller (Titan-like) planetary radius. 

The final two waves shown in Figure \ref{fig:waves2}c and d are additional retrograde modes. The first (Figure \ref{fig:waves2}c) resembles an equatorial Rossby wave, while the second (Figure \ref{fig:waves2}d) appears to be the same mixed Rossby--gravity wave identified in the upper stratosphere. Neither of these waves induces a significant momentum flux near the equator. This is likely due to their lower amplitude close to the equator, and possibly also because their phase speed is far from the mean flow speed.

\begin{figure}
    \centering\includegraphics[width=.95\linewidth]{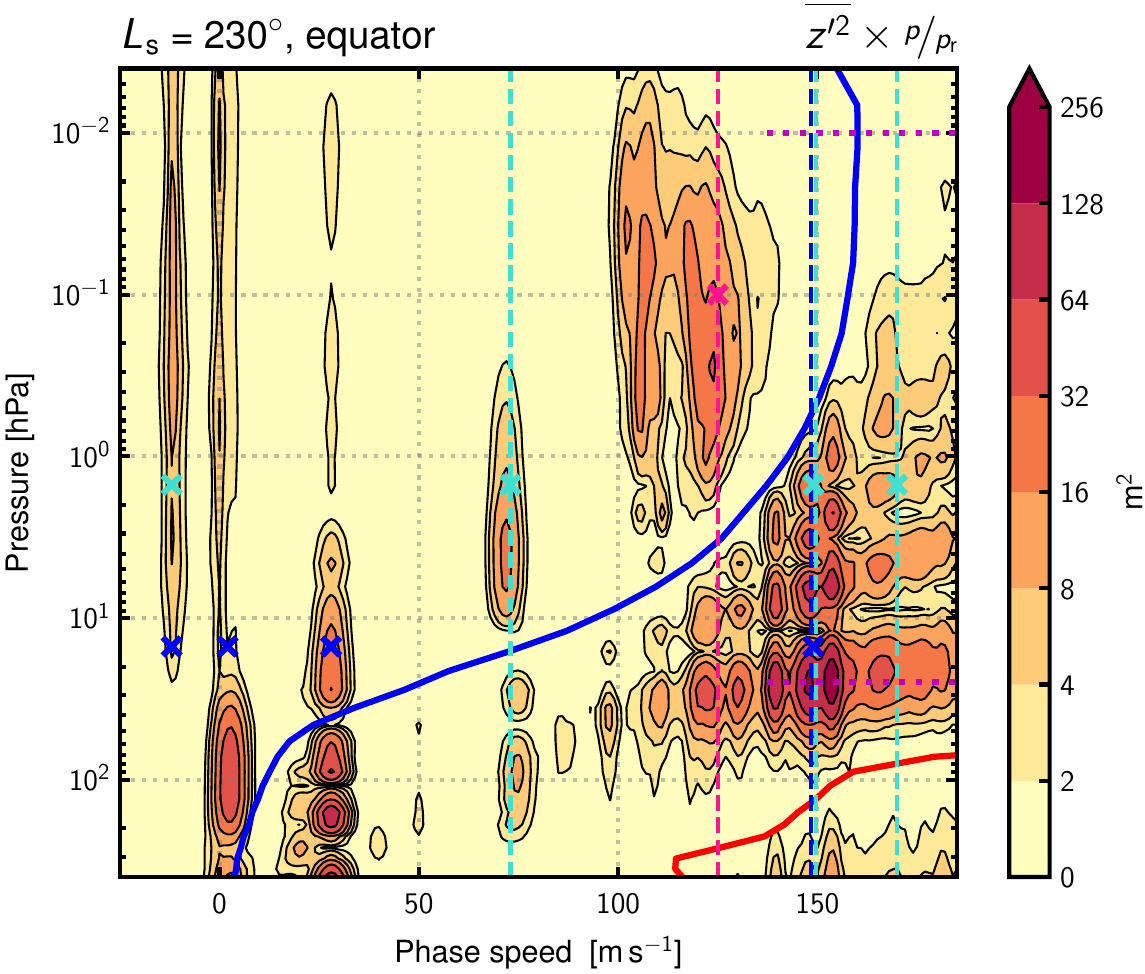}
    \caption{Eddy geopotential height { spectrum} presented as a function of pressure and phase speed, averaged within $\pm10^{\circ}$ latitude { for $L_{\text{s}} = 226^{\circ}$ -- $234^{\circ}$.} Vertical dashed lines correspond to $c_p$ associated with waves whose spatial structure in the $\lambda$--$p$ plane is shown in Figure \ref{fig:waves_vert1} or Figure \ref{fig:waves_newER}. Points marked by cyan, blue and pink crosses correspond to the locations of waves whose horizontal structures are shown in Figures \ref{fig:waves1}, \ref{fig:waves2}, and \ref{fig:waves_newER}, respectively. The solid blue curve shows the zonal-mean zonal wind $\overline{u}$. The red curve indicates the pressure below which waves should be evanescent for a given phase speed (see text). The dotted purple lines indicate the pressure range spanned in Figures \ref{fig:seasonal_TEM} and \ref{fig:EP_eastwest1}.}\label{fig:vert_spec1}
\end{figure}

\subsection{Vertical structure}

\begin{figure}
    \centering\includegraphics[width=0.925\linewidth]{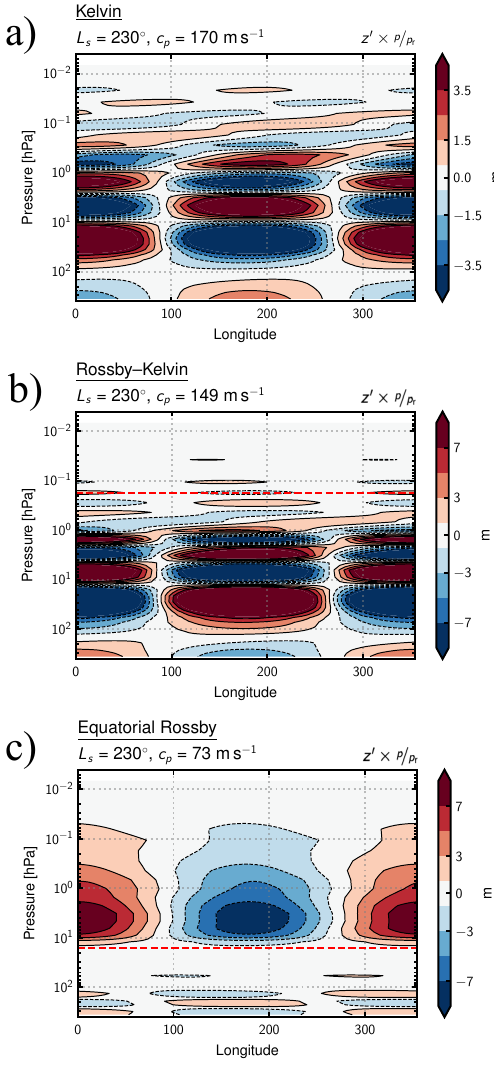}
    \caption{{ Composites showing} filtered geopotential height anomaly averaged between $\pm20^{\circ}$ latitude for selected waves. Red dashed lines in panels b) and c) indicate the pressure where $c_p=\overline{u}$ for the selected phase speed. Above these lines, waves are retrograde, and below they are prograde. As in Figures \ref{fig:waves1} and \ref{fig:waves2}, fields are filtered for frequencies corresponding to a range of phase speeds $c_p=c_{p,\text{wave}}\pm c_{p,\text{diff}}$, with $c_{p,\text{diff}}=5\,\text{m\,s}^{-1}$.}\label{fig:waves_vert1}
\end{figure}

Some of the waves identified in our analysis  of the $p=1.5\,\text{hPa}$ and $p=16.1\,\text{hPa}$ levels appear to have a coherent vertical structure. Notably, we identified a mixed Rossby--gravity wave in both the upper and lower stratosphere with $c_p=-12\,\text{m\,s}^{-1}$. Additionally, the eddy geopotential height spectrum analysed at $p=16.1\,\text{hPa}$ (Figure \ref{fig:spec2}a) has a Kelvin wave-like feature that appears across a broad range of phase speeds, $c_p=125$ -- $200\,\text{m\,s}^{-1}$, similar to those of the Rossby--Kelvin waves identified at $p=1.5\,\text{hPa}$. Using this as a motivation, in this section we analyse the vertical structure of the waves present during the period  $L_{\text{s}} = 226^{\circ}$ -- $234^{\circ}$.

Figure \ref{fig:vert_spec1} shows the pressure--phase speed spectrum of $\overline{z^{\prime2}}$, averaged between $\pm10^{\circ}$ latitude. The background zonal-mean zonal wind is also plotted as a solid blue line. This analysis is presented over a broad range of pressures, from $4\times10^{2}\,\text{hPa}$ to $4\times10^{-3}\,\text{hPa}$. 

\begin{figure*}
    \centering\includegraphics[width=0.9\textwidth]{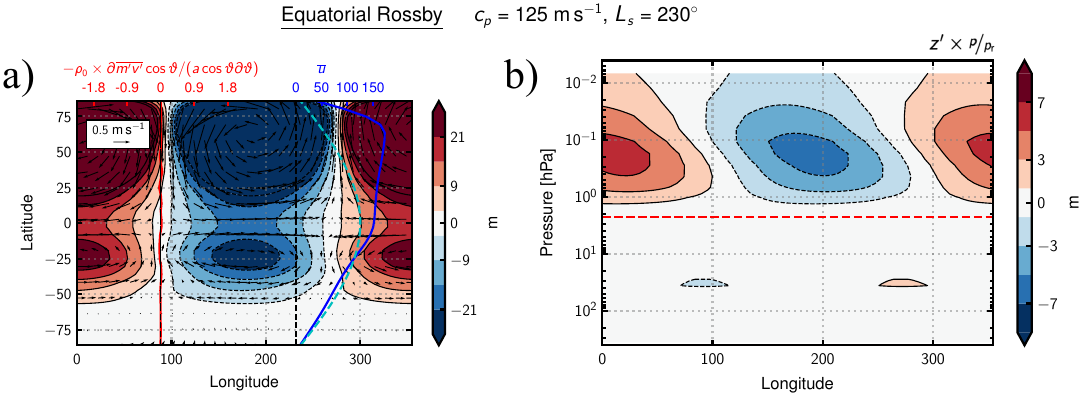}
    \caption{ Composites showing the horizontal and vertical structure of the retrograde disturbance identified in Figure \ref{fig:vert_spec1} at $p=0.1\,\text{hPa}$ and $c_{p}=125\,\text{m\,s}^{-1}$. As with other composites, frequencies were filtered using a window of width $5\,\text{m\,s}^{-1}$ in phase speed space.}\label{fig:waves_newER}
\end{figure*}

A number of disturbances with a coherent vertical structure can be identified. This is particularly evident for retrograde disturbances (with $c_p<\overline{u}$); the signature of the mixed Rossby-gravity wave is present at $-12\,\text{m\,s}^{-1}$ and appears to extend from  roughly $20\,\text{hPa}$ to very high altitude, and each of the equatorial Rossby waves identified in Figures \ref{fig:waves1} and \ref{fig:waves2}, with phase speeds $c_p=2.5,\,28,\ \text{and}\,73\,\text{m\,s}^{-1}$, appear over broad pressure ranges. 

Prograde waves also exist across a broad range of phase speeds and pressures, but with a complicated vertical structure. In the lower stratosphere, the $\overline{z^{\prime2}}$ spectrum appears to be amplified at certain pressures for certain phase speeds. This distribution is difficult to interpret, although the region where it is most prominent notably coincides with sharp vertical contrasts in the background wind $\overline{u}$ and the buoyancy frequency $N$ (see Figure \ref{fig:NT}). This effect is not realised for the retrograde waves, and less so for prograde waves in the upper stratosphere, which both have a more evenly spread $\overline{z^{\prime2}}$ distribution in the vertical. Finally, we note that there is clear evidence for a critical region in Figure \ref{fig:vert_spec1} close to the mean flow speed (i.e., where $\lvert c_p-\overline{u}\rvert$ is small), through which disturbances cannot propagate vertically and the wave amplitudes become small.

Figure \ref{fig:waves_vert1} presents the vertical structure of selected waves in the longitude--pressure plane, averaged between $\pm20^{\circ}$ latitude. These composites are obtained in the same way as those presented in Figures \ref{fig:waves1} and \ref{fig:waves2}. Each wave is indicated by a vertical dashed line in Figure \ref{fig:vert_spec1}, and corresponds to one  whose horizontal structure is shown in Figure \ref{fig:waves1} or \ref{fig:waves2} (or both). In panels b and c, red dashed lines indicate the pressure where $c_p=\overline{u}$. In these panels, disturbances are retrograde above the red dashed line, and prograde below. 

Figures \ref{fig:waves_vert1}a and b show the vertical structure of $\overline{z^{\prime}}$ for the $c_p=170\,\text{m\,s}^{-1}$ and $c_p=149\,\text{m\,s}^{-1}$ Rossby--Kelvin waves shown in Figure \ref{fig:waves1}a and b for $p=1.5\,\text{hPa}$, which more closely resemble pure Kelvin waves in the lower stratosphere. These modes have a high vertical wavenumber, with a wavelength that decreases with altitude. The vertical wavelength is consistent with that selected to produce the  Kelvin wave dispersion curves shown in Figure \ref{fig:spec1}b and d. For linear Kelvin waves on the equatorial beta plane, the intrinsic frequency obeys the dispersion relationship \citep{1987mad..book.....A} \begin{equation}
    \omega^{+}_{\text{kelvin}}=-\frac{Nk}{[m^{2}+1/(4H^{2})]^{\nicefrac{1}{2}}}, \label{eq:K_dispersion}
\end{equation}
where $m$ is the vertical wavenumber, { related to the vertical wavelength $\lambda_{\text{v}}$ via $m=2\pi/\lambda_{\text{v}}$ (see Appendix \ref{sec:appendix} for more detail)}. For waves with smaller $\omega^{+}$, travelling at a speed close to the mean flow speed, we expect larger $m$. Furthermore, for prograde waves propagating vertically in a positive (increasing with altitude) shear, it is necessary for the vertical wavelength to shorten with altitude so as to preserve a coherent $c_p$ in the vertical \citep{1987mad..book.....A,2006JAtS...63.1623I}, { although this effect is weakened slightly by vertical variation in $N$, as smaller $N$ requires smaller $m$ (larger $\lambda_{\text{v}}$) for $\omega$ to remain constant ($N$ decreases with altitude above $p=10\,\text{hPa}$; see Figure \ref{fig:NT}). }

Above roughly $p\approx2\,{-}\,3\,\text{hPa}$, these modes have a clear phase tilt with altitude which, for a prograde Kelvin wave, is consistent with upward vertical propagation \citep[see, e.g., Figure 4.19 of][]{1987mad..book.....A}, and suggestive of a source region for the wave around $p=3\,\text{hPa}$. Possible sources of instability that lead to the generation of this wave, and others analysed later, will be discussed in Section \ref{sec:instability}.

Below $p\approx2$\,--\,$3\,\text{hPa}$, the wave appears to take the form of a standing wave that does not propagate vertically, where instead we might expect it to propagate downward away from the apparent source region. This standing structure is coincident with the vertical range in Figure \ref{fig:vert_spec1} where $\overline{z^{\prime2}}$ is concentrated at particular pressures. This behaviour is difficult to interpret using the linear theory, within which equatorial Kelvin waves only become vertically trapped when $N^{2}k^{2}/\omega^{+2}<1/(4H^{2})$, so that $m$ in equation \eqref{eq:K_dispersion} is imaginary \citep{1990JAtS...47..293D}. This is not satisfied in our simulation  around $p\approx3\,\text{hPa}$, but is satisfied much deeper in the atmosphere (below the red curve in Figure \ref{fig:vert_spec1}, which, notably, is roughly where $\overline{z^{\prime2}}$ goes to zero). \citet{1975wavesbook.....G} show that the curvature of the background wind can also inhibit vertical propagation, if $N^{2}k/\lvert\omega^{+}\rvert<\lvert\text{d}^{2}\overline{u}/\text{d}z^{2}\rvert$, but this condition is not satisfied anywhere in our model atmosphere. Possible explanations for apparent disagreement between the linear theory and the standing structure of the Kelvin waves we have identified will be discussed further in Section \ref{sec:vertprop_discuss}.

Figure \ref{fig:waves_vert1}c shows the vertical structure of the equatorial Rossby wave presented in Figure \ref{fig:waves1}c. This wave has a barotropic structure, with a slight phase tilt indicative of upward propagation (for a retrograde wave; cf. \citealp{2011JAtS...68..839Y}) away from a source around $p=15\,\text{hPa}$. This wave has a much longer vertical wavelength than the Rossby--Kelvin/Kelvin waves, consistent with larger separation between $c_p$ and $\overline{u}$.

{ Finally, we note that the $\overline{z^{\prime2}}$ spectrum presented in Figure \ref{fig:vert_spec1} shows that further retrograde disturbances exist at lower pressures in the upper stratosphere than those analysed in Section \ref{sec:upper}. Figure \ref{fig:waves_newER} shows the horizontal and vertical structure of one such wave, located at $p=0.1\,\text{hPa}$ with phase speed $c_{p}=125\,\text{m\,s}^{-1}$. The horizontal structure of this wave resembles that of an equatorial Rossby wave, distorted by the horizontal shear in the background wind. The wave has a barotropic structure in the vertical, with a phase tilt indicative of upward propagation away from a source region located around $p\approx1\,\text{hPa}$. As with the other retrograde waves in the upper stratosphere, this wave does not appear to induce significant angular momentum flux convergence at the equator.}

{

\subsection{Differences between autumn and winter}

We have only presented an analysis of the waves present during northern hemisphere autumn, the period during which eddy angular momentum fluxes are largest (see Figure \ref{fig:seasonal_TEM}). Closer to the solstices (i.e., `summer' or `winter'), the distribution of eddy accelerations acting on the mean flow is quite different; notably, acceleration of the mean flow by waves is only significant in the lower stratosphere (top and bottom rows, Figure \ref{fig:seasonal_TEM}), which implies that a different spectrum of waves is active compared with autumn. 

We also conducted an analysis similar to that presented in the preceding subsections, but for the periods $L_{\text{s}} = 186^{\circ}$ -- $194^{\circ}$ and $L_{\text{s}} = 346^{\circ}$ -- $354^{\circ}$. In the upper stratosphere, only Kelvin waves which are weakly phase-locked with a mid-latitude Rossby component, or absent a Rossby component entirely, were identified (very similar to that shown in Figure \ref{fig:waves1}a). As such, these prograde waves are unable to accelerate the mean flow in the upper stratosphere, consistent with Figure \ref{fig:seasonal_TEM}. 

In the lower troposphere, the results of our analysis were complicated and difficult to interpret. A range of disturbances that contribute to equatorial angular momentum flux convergence were identified, including equatorial Rossby waves and Rossby--Kelvin waves. The waves present are of varying spatial scales (zonal wavenumbers $1$\,--\,$3$ are roughly equally represented) and typically very short vertical wavelengths. However, we found our analysis to be highly sensitive to the time period and pressure level chosen for study. We believe this is due to a combination of: i) the relevant waves' very short vertical wavelengths, ii) strong vertical shear in the background zonal wind, and iii) the relatively coarse vertical resolution used in the model (discussed further in Section \ref{sec:vertprop_discuss}). For these reasons, we choose to not include a full presentation of this analysis in the present study, and instead  propose that investigation of these issues should be a priority for future research. }

\section{Discussion}\label{sec:discuss}

{ 
\subsection{Wave generation mechanisms}\label{sec:instability}
}

In this subsection, { we discuss} possible mechanisms for the generation of the waves we have identified in our simulation. We consider both i) ageostrophic, barotropic instabilities, similar to that discussed by \citet{2005JAtS...62.2514I} and \citet{2014GeoRL..41.4118W} 
, as well as ii) conventional quasigeostrophic barotropic instability \citep[see, e.g.,][]{2020NPGeo..27..147R}.

\subsubsection{Ageostrophic, barotropic instabilities}\label{sec:shear}

In our model, the acceleration of superrotation in the upper stratosphere is due to waves that consist of an equatorial Kelvin-like component, coupled with a Rossby-like component in mid- and high latitudes. This mode grows when there is strong meridional shear in the horizontal wind, which is most pronounced around early northern hemisphere autumn. It is generally asymmetric with respect to the equator, with a Rossby-like component located in the winter hemisphere but not the summer hemisphere, due to asymmetry in the background wind.

Previous work has shown that coupled Rossby--Kelvin waves similar to those identified here can grow due to an ageostrophic, barotropic instability that exists on slowly rotating and/or small-sized planets if there is meridional shear in the background wind (such that the velocity in mid-latitudes is greater than that at the equator; \citealp{2005JAtS...62.2514I,2014GeoRL..41.4118W}). The instability requires the deformation radius to be large enough that the Rossby and Kelvin components have some spatial overlap, and that the two component waves have similar frequencies. 

The first condition is generally true in the slowly rotating/small planet regime as both the equatorial and mid-latitude deformation radii depend inversely on the planetary rotation rate and radius. The second `frequency matching' condition can be quantified via the definition of a `Froude number' that measures the ratio of the Doppler shifted frequency of mid- or high-latitude Rossby waves to that of the equatorial Kelvin wave \citep{2014GeoRL..41.4118W}:
\begin{equation}
    \text{Fr}\equiv\frac{\omega_{\text{rossby}}}{\omega_{\text{kelvin}}}\approx\frac{\overline{u}_{\text{ml}}/\cos\vartheta_{\text{ml}}}{\overline{u}_{\text{eq}} + \omega^{+}_{\text{kelvin}}/k}, \label{eq:Fr}
\end{equation}
where $\omega^{+}_{\text{kelvin}}/k$ is given by equation \eqref{eq:K_dispersion}, $\overline{u}_{\text{eq}}$ is the zonal velocity at the equator, $\overline{u}_{\text{ml}}$ is the mid-latitude zonal wind speed at latitude $\vartheta_{\text{ml}}$, and the intrinsic frequency of the Rossby component has been assumed small and neglected. In principle, $\text{Fr}=1$ for a Rossby--Kelvin mode. In practise, however, \citet{2014GeoRL..41.4118W} find that neglecting the intrinsic component of the Rossby wave frequency means that Rossby--Kelvin instability occurs for a range of $\text{Fr}>1$. Specifically, they find the growth rate (and equatorward eddy momentum flux due to the wave) is maximised when $\text{Fr}\approx2$, and drops off as $\text{Fr}$ is varied away from $2$ (but within the range $\text{Fr}=1$\,--\,$3$, outside of which the instability is inactive).

\begin{figure}
    \centering\includegraphics[width=.925\linewidth]{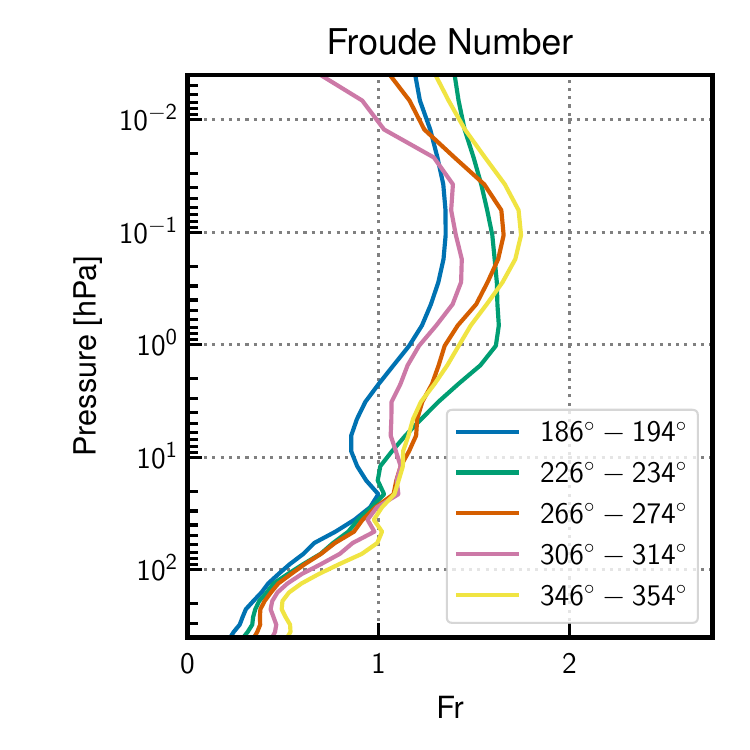}
    \caption{Froude number as in \citet{2014GeoRL..41.4118W}, indicating the ratio of the Kelvin wave frequency to the Rossby wave frequency (see equation \ref{eq:Fr} and accompanying text for details). Lines are shown for different { ranges of $L_{\text{s}}$}, indicated in the legend. }\label{fig:Fr}
\end{figure}

Figure \ref{fig:Fr} shows $\text{Fr}$ as a function of pressure at different times during the Titan year (corresponding to those shown in Figure \ref{fig:seasonal_TEM}). { For each line, we have set the vertical wavelength $\lambda_{\text{v}}=30\,\text{km}$ for the computation of $\omega^{+}_{\text{kelvin}}$ (see Appendix \ref{sec:appendix}), chosen to be characteristic of the vertical wavelength of the prograde waves identified in the simulation (i.e., Figure \ref{fig:waves_vert1}a,b)}, and we fix $\vartheta_{\text{ml}}=50^{\circ}\,\text{N}$ (our computation is not senstive to this choice).  { For each period shown, $1\lesssim\text{Fr}<2.5$ throughout the depth of the stratosphere, indicating conditions favorable for Rossby--Kelvin instability. Notably, at pressures $p\approx1$\,--\,$3\,\text{hPa}$ (from which Rossby--Kelvin waves appear to originate), $\text{Fr}$ is largest during the period when coupled Rossby--Kelvin modes induce the greatest acceleration, as in \citet{2014GeoRL..41.4118W}. 

Although our analysis is not fully conclusive regarding the origin of the waves responsible for accelerating superrotation in the upper stratosphere}, it is at least \emph{consistent} with the waves being generated by an ageostropic shear instability that is most active during autumn, when the meridional shear in the background wind is maximised. 

\begin{figure}
    \centering\includegraphics[width=.975\linewidth]{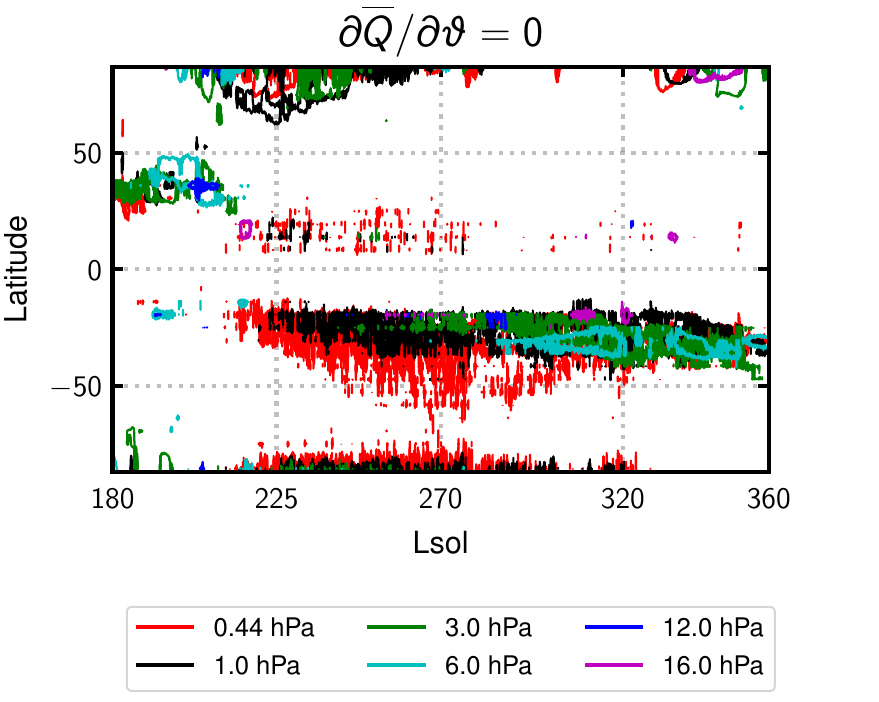}
    \caption{Criterion for barotropic instability plotted as zero contours for $\partial\overline{Q}/\partial\vartheta\rvert_\theta$ (see text). Contours are shown for a range of pressures indicated in the legend.} \label{fig:baro}
\end{figure}

\subsubsection{Conventional barotropic instability}\label{sec:barotropic}

In the lower stratosphere, equatorial eddy angular momentum flux convergence displays less seasonal variation (Figure \ref{fig:seasonal_TEM}). For at least part of the year, this convergence is due to retrograde equatorial Rossby waves, and, more generally, retrograde modes exist throughout the year in the stratosphere. This suggests a generation mechanism for retrograde waves that does not depend as strongly with the seasonal cycle as the ageostrophic, barotropic instability described in the previous subsection. A possible mechanism is conventional quasigeostrophic barotropic instability, which has been shown to occur close to the equator in other Titan GCMs \citep{2011Icar..213..636N,2012Icar..218..707L}.

A necessary criterion for quasigeostrophic barotropic instability is that the meridional gradient of zonally-averaged potential vorticity, $\overline{Q}$, taken on isentropic surfaces, change sign \citep{1962JAtS...19..159C}. This requires that  \begin{equation}
    \xi\equiv\partial \overline{Q}/\partial\vartheta\rvert_\theta=0
\end{equation}
somewhere within the domain. $Q$ is defined \begin{equation}
    Q = -g(f+\mathbf{k}\cdot\nabla_\theta\times\boldsymbol{v})/(\partial p/\partial \theta),
\end{equation} 
where $\nabla_\theta$ is the horizontal gradient acting along surfaces of constant potential temperature, $\theta$ \citep{1985QJRMS.111..877H}.

Figure \ref{fig:baro} shows zero-contours for $\xi$ at a range of pressures in the stratosphere, over the period { $L_{\text{s}} = 180^{\circ}$ -- $360^{\circ}$}. The criterion for barotropic instability is satisfied year-round at pressures from which the retrograde waves shown in Figures \ref{fig:waves_vert1} and \ref{fig:waves_newER} appear to originate. $\xi=0$ generally occurs at latitudes conincident with the equatorward flank of the summer hemisphere mid-latitude jet. We note that $\xi=0$ never occurs directly on the equator. However, the occurence of barotropic instability within a deformation radius of the equator is sufficient to generate equatorial wave modes (as is the case here). 

The fact that $\xi=0$ for most of the year in regions where retrograde equatorial waves appear to be generated points to barotropic instability as a likely source of excitation. However, as before, we cannot definitively assert that this is the case based solely on our analysis. Instead, this suggestion, and the suggestion that ageostrophic barotropic instability leads to the generation of Rossby--Kelvin type waves in the upper stratosphere, should be investigated further by conducting stability analyses of mean circulation states that are characteristic of Titan.

\subsection{Relationship to results from idealised modeling studies}

As described in the introduction, the generation and maintenance of superrotation in the atmospheres of small and/or slowly rotating planets has been investigated extensively with idealised `Earth-like' GCMs forced by time-independent, hemispherically symmetric heating and cooling. However, it is not obvious \emph{a priori} that mechanisms leading to the generation and maintainance of superrotation in a simplified modeling setting are of direct relevance to Titan's atmosphere, which is complicated by seasonally-varying, hemispherically asymmetric meridional shear. 

Studies with idealised GCMs have shown that a global mode that couples equatorial Kelvin waves with Rossby-like waves in midlatitudes leads to the generation of superrotation during the model spin-up period \citep{2010JGRE..11512008M}. Phase locking of the Rossby and Kelvin components is possible as the Rossby waves are Doppler shifted eastwards by a midlatitude jet that emerges very early in the spin-up phase (and is associated with conservation of angular momentum within the Hadley cell). However, once the Rossby--Kelvin wave has accelerated the zonal wind at the equator and the simulation equilibrates, the Kelvin wave also experiences a Doppler shift (due to the equatorial jet) and phase locking between the Rossby and Kelvin components is no longer possible. \citet{2014Icar..238...93D,2016JAtS...73.3181D} have shown that in equilibrium superrotation is maintained against dissipation by equatorial Rossby waves, which develop a meridional phase tilt in vertical shear that facilitates eddy momentum flux convergence \citep{2006JAtS...63.1623I}. 

Our analysis suggests that direct analogies between the generation scenario \emph{and} the maintenance scenario may be drawn with different aspects of Titan's stratospheric circulation. The seasonal maintenance of superrotation by a Rossby--Kelvin like wave in the upper stratosphere is analogous to the \emph{generation} of superrotation by Rossby-Kelvin waves in idealised Earth-like GCMs during their spin-up phase. In Titan's upper stratosphere, the circulation in each hemisphere never really equilibrates, and is instead constantly `spinning up' and `spinning down' with the seasonal cycle. The maintenance mechanism for superrotation in Titan's lower stratosphere appears to be more complicated. During northern hemisphere autumn, equatorial Rossby waves play a central role in maintaining superrotation, similar to the \emph{maintenance} mechanism identified in idealised GCMs by \citet{2014Icar..238...93D,2016JAtS...73.3181D}. { However, our analysis of other periods during the seasonal cycle (not shown) indicates this is not always the case. In the lower stratosphere, strong vertical shear, and weak (but non-negligible) hemispheric asymmetry in Titan's zonal wind differentiates the circulation from that obtained in idealised models, and can have an important influence on the types of waves that maintain superrotation. }

\citet{2014ApJ...787...23M} have shown that when a seasonal cycle is included in an idealised Earth-like model, the generation of equatorial superrotation is inhibited. They suggest this is due to two effects: i) cross-equatorial advection of momentum by the overturning circulation, and ii) suppression of the Rossby--Kelvin instability in hemispherically asymmetric meridional shear. To achieve a Titan-like superrotating circulation, they increase the radiative cooling time in their model by a factor of 100 (appropriate for Titan), the effect of which is to reduce the strength of  overturning which, in turn, damps each of the superrotation-inhibiting seasonal effects described above. 

In our simulation, \emph{there is} strong meridional shear, but this actually appears to give rise to the asymmetric Rossby--Kelvin type wave that accelerates superrotation in the upper stratosphere (notably, this disturbance is very similar to one shown in \citealp{2014ApJ...787...23M} for a Titan-like case that does not develop strong superrotation; compare our Figure \ref{fig:waves1}b with their Figure 8). This indicates that Rossby--Kelvin modes that are not symmetric about the equator can provide sufficient momentum transport to maintain superrotation, counter to the suggestion of \citet{2014ApJ...787...23M}. 
However, it is likely that the superrotation in our simulation does benefit from effect ii) above, namely `reduced mean flow momentum transport due to weaker overturning', as a result of Titan's long radiative timescale, although this will be reduced in the stratosphere where the density is low. Additionally, we note that in our simulation, the overturning in the stratosphere is partially detached from that in the troposphere, which limits the extent to which the mean flow can communicate the effects of friction near the surface to the superrotating jet (see, e.g., Figure \ref{fig:zm_fulldepth}, where it could be argued that overturning in the stratosphere takes the form of an isolated cell sitting above another overturning cell in the troposphere).

\subsection{Comparison with other Titan GCMs}

In this subsection we compare our results with those of \citet{2011Icar..213..636N} and \citet{2012Icar..218..707L}, who analysed the waves responsible for accelerating superrotation in TitanWRF and the IPSL Titan GCM, respectively. 

Each of these studies found that eddy angular momentum transport is strongest in the autumn/winter hemisphere, and induces acceleration close to the equator paired with deceleration in mid-latitudes, which is consistent with the seasonal cycle of EP flux divergence presented here in Figure \ref{fig:seasonal_TEM}. 

Spectral analysis of the waves present in the IPSL model also appears to be consistent with some of the results presented here. In particular, the acceleration of superrotation above $p=10\,\text{hPa}$ in their model seems to be due to modes that are prograde at the equator, and, notably, have a similar spatial structure in latitude and longitude to the hemispherically asymmetric Rossby--Kelvin-wave we have identified here (compare the filtered $u^{\prime}$ in \citealp{2012Icar..218..707L}'s Figure 16b with the zonal orientation of quivers in our Figure \ref{fig:waves1}b). A few modes have a coherent structure in the vertical direction in their Figure 14, which shows $u^{\prime}$ and $v^{\prime}$ as a function of frequency and pressure (similar to Figure \ref{fig:waves_vert1} in this work), although not as obviously as in our analysis. There appear to be some retrograde modes in the lower stratosphere of the IPSL simulation, but \citet{2012Icar..218..707L} do not show whether or not they contribute to the angular momentum budget at the equator. 

\citet{2011Icar..213..636N} present latitude--phase speed eddy momentum flux co-spectra for TitanWRF, which show that acceleration in TitanWRF's upper stratosphere is due to a mode with strong coherence along a line of $c_p\times\cos\vartheta$, which has $c_p>\overline{u}$ at the equator and $c_p<\overline{u}$ in mid-latitudes. They interpret this in terms of a wave propagating meridionally from the equator and breaking in mid-latitudes after crossing a critical line where $c_p=\overline{u}$, whereas our analysis indicates that this sort of signature is due to a coupled Rossby--Kelvin mode that does not propagate meridionally. However, \citet{2011Icar..213..636N} do not present the horizontal latitude--longitude structure of the wave that accelerates superrotation in the upper stratosphere of TitanWRF, so it is difficult to comment on whether or not it is similar to those identified in this study. 

We note that \citet{2011Icar..213..636N} present evidence that angular momentum transport from the winter hemisphere to the equator is concentrated into short-lived  `transfer events' (duration { corresponding to $<5^{\circ}\,L_{\text{s}}$}). While there is eddy angular momentum flux variability in our simulation that occurs on a { timescale corresponding to a few degrees $L_{\text{s}}$}, we have not identified any events of comparable magnitude to those identified in TitanWRF. \citet{2011Icar..213..636N} show that the timing of these transfer events is closely related to instances when barotropic instability might operate (i.e., times when $\xi=0$), which also occurs in short-lived events { (note that in our model we do not identify this sort of behaviour, cf. Figure \ref{fig:baro}, where $\xi=0$ occurs continuously throughout the year, as opposed to occurring in short bursts)}. This might indicate that the waves that accelerate superrotation in the upper stratosphere of TitanWRF are generated by a different mechanism to those here.

\subsection{Vertical structure of equatorial Kelvin waves}\label{sec:vertprop_discuss} 

In this study, we have identified a number of waves which resemble an equatorial Kelvin wave at the equator (see, e.g., Figure \ref{fig:waves1}a,b and Figure \ref{fig:waves2}a). These waves generally have short vertical wavelengths, consistent with their phase speeds being close to the mean flow speed (cf. equation \ref{eq:K_dispersion}). Furthermore, there is evidence that { their wavelengths vary} due to vertical propagation in a vertically-sheared background zonal wind profile (e.g., Figure \ref{fig:waves_vert1}a), which, again, is consistent with our expectation from linear theory. 

However, the equatorial Kelvin waves we have identified also exhibit some unusual characteristics. Specifically, a number of these waves appear to be vertically trapped (i.e., unable to propagate) in the region where the background vertical wind shear is strongest (Figures \ref{fig:waves_vert1}a,b). This behaviour is not predicted by the linear theory for waves on the equatorial beta plane. However, the linear theory is only valid for the case of a slowly varying background $\overline{u}$ and $N$ (see, e.g., Appendix 4A in \citealp{1987mad..book.....A}), which is clearly not satisfied in the stratosphere of our Titan GCM. It is therefore \emph{possible} that unusual vertically trapped behaviour exhibited by the Kelvin waves is a physical response to the strong vertical variation in $N$ and $\overline{u}$ in our simulation. { For example, it may be the case that upward propagating Kelvin waves are reflected when they encounter the critical region where $\lvert c_{p}-\overline{u}\rvert$ is small \citep{1972JGR....77.2915M,2008JASTP..70..742W}, or that downward propagating waves are reflected upwards at the boundary where they are expected to become evanescent (i.e., the red line in Figure \ref{fig:vert_spec1}; \citealp{1997JGR...10226301I,2008JGRD..11324104P}), and the resulting superposition of upward and downward propagating waves creates a standing structure. Furthermore, the possibility of reflection at the $c_{p}=\overline{u}$ critical layer appears to be consistent with the amplification of the $\overline{z^{\prime2}}$ spectrum shown in Figure \ref{fig:vert_spec1} for disturbances `below' the zonal-mean wind line, compared with those for which $c_{p}$ is greater than $\overline{u}$ at all pressures.} \citet{2006JAtS...63.1623I} investigated the properties of equatorial waves on Venus using a non-linear model of Venus' stratosphere with a prescribed background state. We suggest that a similar study using a Titan-like background, to ascertain the `expected' properties of equatorial waves in Titan's stratosphere, is an important topic for future work motivated by the present study. 

Notwithstanding the possible influence of strong vertical variation in $\overline{u}$ and $N$, it is likely that the short vertical wavelength of the equatorial waves in our simulation means that they are not properly resolved with the model's current vertical resolution. At least four grid points per wavelength are required for a wave to be represented accurately, i.e., $\lambda_{\text{v}}/(4\,\delta z^{\ast})>1$, where $\delta z^{\ast}$ is the vertical grid spacing. This is not always satisfied in our simulation; for example, $\lambda_{\text{v}}/(4\,\delta z^{\ast})\approx0.9$ for the Rossby--Kelvin / Kelvin wave shown in Figure \ref{fig:waves_vert1}a (example uses $\lambda_{\text{v}}=22\,\text{km}$ and $\delta z^{\ast}=6\,\text{km}$, which are representative for $p\approx2\,\text{hPa}$).
\citet{1992JAtS...49..785B} investigated the effect of vertical resolution on equatorial waves in the stratosphere of a GCM designed for Earth. They found that for $\lambda_{\text{v}}/(4\,\delta z^{\ast})\lesssim1$, the amplitude of equatorial waves depended strongly on the vertical resolution (increasing amplitude for increasing resolution). This, in turn, affects eddy acceleration of the mean flow, as the EP flux is formed from covariances of wave properties, and so depends on the wave amplitude squared. \citet{1992JAtS...49..785B} also note that insufficient resolution in regions with strong vertical shear in the background wind can lead to poor representation of vertically propagating equatorial waves; notably, it is in such regions that the equatorial Kelvin waves we identify exhibit unusual behaviour.

It is therefore possible that the EP flux, and its divergence, that are obtained from our current simulation of Titan using TAM only partially capture the way in which waves interact with the mean flow in Titan's real atmosphere. The vertical resolution used for this study is similar to that of other Titan GCMs (e.g., TitanWRF and the IPSL Titan Model; \citealp{2011Icar..213..636N}; \citealp{2012Icar..218..707L}), and thus we expect that the issues regarding vertical resolution discussed here for TAM extend to those models as well. Poor representation of the vertical propagation and structure of equatorial waves may explain why current Titan GCMs struggle to produce the `zonal wind minimum' present in the Huygens descent profile (at $p=20\,\text{hPa}$ in Figure 2 of \citealp{2005Natur.438..800B}), which could be formed by a vertical eddy momentum flux divergence (e.g., due to prograde waves propagating vertically away from this region). We believe that our results motivate the development of Titan GCMs with higher vertical resolution in the stratosphere.

\subsection{Similarities between equatorial waves in Titan's troposphere and stratosphere}

Our analysis has shown that a range of equatorial waves, including Kelvin, equatorial Rossby, and mixed Rossby--gravity waves can exist in conditions characteristic of Titan's stratosphere. We briefly note here that similar equatorial modes have also been identified in the troposphere of TAM simulations \citep{2021GeoRL..4894244B}, as have high-latitude Rossby-like disturbances \citep[also identified in our analysis of the stratosphere;][]{2015GeoRL..42.6213L,2021GeoRL..4894244B}. Equatorial Kelvin waves and mid-latitude Rossby waves have also been inferred from observations of clouds in Titan's troposphere \citep{2009Natur.460..873S,2011NatGe...4..589M}.

A number of the equatorial waves we have identified in this work have vertical structures consistent with vertical group propagation. Likewise, we would expect equatorial waves in Titan's troposphere to propagate vertically as well as horizontally. This raises the possibility that waves generated in Titan's troposphere could interact with those in the stratosphere, or vice versa. There is some evidence for this in the pressure--phase speed diagram shown in Figure \ref{fig:vert_spec1}, where prograde waves in the upper troposphere (where $\overline{u}$ is smaller, so $c_p-\overline{u}>0$) and retrograde modes in the stratosphere (where $\overline{u}$ is greater, so $c_p-\overline{u}<0$) exist at the same phase speeds (e.g., $c_p=28\,\text{m\,s}^{-1}$ in Figure \ref{fig:vert_spec1}). It is possible, for example, that prograde waves which propagate upwards from the upper troposphere and break when $c_p=\overline{u}$ could provide the disturbances from which the retrograde waves in the stratosphere grow. An analysis of the possible interactions between waves in Titan's troposphere and stratosphere is beyond the scope of this work, but should be pursued in the future. 

\vspace*{.1in}

\section{Conclusions}\label{sec:conclude}

The aim of this work was to characterise the different wave modes that are supported in a simulation of Titan's stratosphere using the Titan Atmospheric Model \citep{2015Icar..250..516L,2023Icar..39015291L}, with a focus on those that contribute most to the acceleration of equatorial superrotation. 

The key conclusions of this study are as follows: 
\begin{enumerate}
    \item A variety of equatorial waves are identified in TAM's stratosphere, including equatorial Kelvin waves, equatorial Rossby waves, and mixed Rossby--Gravity waves. 
    \item In the upper stratosphere, equatorial waves are strongly affected by the seasonal cycle and feature pronounced hemispheric asymmetries when meridional shear in the background wind is strongest. This leads to interactions between equatorial Kelvin waves and Rossby-like waves in the winter hemisphere, which converge momentum from winter mid-latitudes towards the equator, opposing momentum transport away from the equator by the mean flow. 
    \item In the lower stratosphere, equatorial waves have more hemispheric symmetry, and equatorial superrotation is accelerated throughout the year. Equatorial Rossby waves play an important role in inducing equatorial eddy angular momentum flux convergence in this region. 
    \item Seasonal variation in the Rossby--Kelvin type waves that accelerate superrotation in the upper stratosphere is consistent with generation by ageostrophic, barotropic `Rossby--Kelvin instability' \citep[cf.][]{2014GeoRL..41.4118W}. Meanwhile, conventional quasigeostrophic barotropic instability is a plausible source of waves that accelerate superrotation in the lower stratosphere. A definitive analysis of the instabilities responsible for exciting the waves we have identified is beyond the scope of the present work. 
    \item Equatorial waves that are most important for accelerating superrotation generally have phase speeds that are close to the zonal velocity of the mean flow (i.e., small intrinsic frequency $\omega^{+}$). Small $\omega^{+}$ requires these modes to have short vertical wavelengths, and, therefore, it is likely that they are underresolved in the vertical. 
\end{enumerate}

In light of our conclusions (in particular conclusions 4 and 5), we suggest that Titan modelers should work to increase the vertical resolution of Titan GCMs. This should be complemented by theoretical investigation of the expected properties of equatorial waves in Titan-like background conditions with strong vertical variation in the background wind and buoyancy frequency, and hemispherically asymmetric meridional shear. 

\vspace{.2in}

\noindent We are grateful for the comments of two anonymous referees that helped us to refine this work throughout. Neil Lewis and Peter Read were supported by the Science and Technology Facilities Council under grant agreements ST/S505638/1, ST/S000461/1, and ST/N00082X/1. Juan Lora and Nick Lombardo acknowledge funding from NASA Cassini Data Analysis Program Grant  80NSSC20K0483.

\appendix

\section{Dispersion curves and deformation radii for equatorial waves}\label{sec:appendix}

In this appendix we briefly describe the method used to compute the dispersion curves plotted in Figure \ref{fig:spec1}, and deformation radii discussed in the text, for various equatorial waves. We follow \citet{WHEELER2015102}, and the reader is directed there for a more thorough discussion of the theory for linear waves on the equatorial beta plane. 

The dispersion curves shown in Figure \ref{fig:spec1} can be obtained by seeking wave-like solutions of the form \begin{equation}
    (u,v,\Phi) = \text{Re}\left\lbrace[\hat{u}(y),\hat{v}(y),\hat{\Phi}(y)]\exp[i(kx-\omega t)]\right\rbrace \label{eq:wave_sol}
\end{equation}
to the linearised shallow water equations on the equatorial beta plane: \begin{align}
    \frac{\partial u_{l}}{\partial t} - \beta yv_{l} &= -\frac{\partial\Phi_{l}}{\partial x}, \label{eq:sw1}\\ 
    \frac{\partial v_{l}}{\partial t} + \beta yu_{l} &= -\frac{\partial\Phi_{l}}{\partial y}, \\ 
    \frac{\partial \Phi_{l}}{\partial t} + gh_{l}\left(\frac{\partial u_{l}}{\partial x_{l}} + \frac{\partial v_{l}}{\partial y}\right) &= 0, \label{eq:sw3}
\end{align}
where $\beta=2\Omega/a$, $u$ and $v$ are the velocity in the $x$ and $y$ directions, respectively, $\Phi=gz$ is the geopotential, and $h_{l}$ is the equivalent depth. $k$ is the zonal wavenumber, $\omega$ is frequency, and $\hat{u}$, $\hat{v}$ and $\hat{\Phi}$ are meridional structure functions. The subscript $l$ indicates that these equations only govern the horizontal structure of a particular vertical normal mode. The vertical structure of each equatorial wave mode is obtained by multiplying the shallow water solutions by the vertical structure function $G_l(z)$, that is \begin{equation}
    u(x,y,z,t) = u_{l}(x,y,t)\times G_{l}(z), 
\end{equation}
where \begin{equation}
    G_{l} = \exp[z/(2H)]\exp(\pm imz),
\end{equation}
and $m$ is the vertical wavenumber, which is related to the shallow water equivalent depth $h_{l}$ via the relation \begin{equation}
    \sqrt{gh_{l}} = \frac{N}{[m^{2}+1/(4H^{2})]^{\nicefrac{1}{2}}}.\label{eq:depth}
\end{equation}

Substitution of equation \eqref{eq:wave_sol} into equations \eqref{eq:sw1}\,--\,\eqref{eq:sw3} yields the general dispersion relation \begin{equation}
    \frac{\sqrt{gh_{l}}}{\beta}\left(\frac{\omega^{2}}{gh_{l}}-k^{2}-\frac{k}{\omega}\beta\right) = 2n+1;\quad n=0,1,2,\dots \label{eq:gen_disp}
\end{equation}
where $n$ is a positive integer called the meridional mode number. Approximate solutions to equation \eqref{eq:gen_disp} may be obtained by considering the limits where $\omega\gg1$ and $\omega\ll1$. If $n\ge1$, then the high frequency approximation to equation \eqref{eq:gen_disp} is \begin{equation}
    \omega_{\text{IG}}\approx\pm\left[(2n+1)\beta\sqrt{gh_{l}}+k^{2}gh_{l}\right]^{\nicefrac{1}{2}}, \label{eq:disp_IG}
\end{equation}
and the low frequency approximation is \begin{equation}
    \omega_{\text{ER}}\approx-\frac{\beta k}{k^{2}+(2n+1)\beta/\sqrt{gh_{l}}}. \label{eq:disp_ER}
\end{equation}
Equation \eqref{eq:disp_IG} has two roots corresponding to eastward (positive root) and westward (negative root) inertio--gravity waves, respectively, and equation \eqref{eq:disp_ER} has one root corresponding to westward propagating equatorial Rossby waves. For $n=0$, an additional (exact) solution to equation \eqref{eq:gen_disp} exists, \begin{equation}
    \omega_{n=0} = k\sqrt{gh_{l}}\left[\frac{1}{2}\pm\frac{1}{2}\left(1+\frac{4\beta}{k^{2}\sqrt{gh_{l}}}\right)^{\nicefrac{1}{2}}\right]. \label{eq:disp_MRG}
\end{equation}
The two roots of equation \eqref{eq:disp_MRG} correspond to an eastward inertio--gravity wave (positive root) and a mixed Rossby--gravity wave (negative root). Finally, an additional wave-like solution exists to equations \eqref{eq:sw1}\,--\,\eqref{eq:sw3} if $\hat{v}=0$ is set, namely \begin{equation}
    \omega_{\text{K}}=\sqrt{gh_{l}}k, \label{eq:disp_K}
\end{equation}
which corresponds to an eastward propagating Kelvin wave. 

We obtain each of the dispersion curves shown in Figure \ref{fig:spec1} as follows. For the Kelvin wave, and the mixed Rossby-gravity and $n=0$ eastward inertio--gravity waves, $\omega$ is computed directly from equations \eqref{eq:disp_K} and \eqref{eq:disp_MRG}, respectively. For $n\ge1$ waves, dispersion curves are obtained by solving equation \eqref{eq:gen_disp} numerically, using the approximate relations given by equations \eqref{eq:disp_IG} and \eqref{eq:disp_ER} as initial guesses. 

{ Dispersion curves are shown in Figure \ref{fig:spec1} for vertical wavelengths $\lambda_{\text{v}}=400\,\text{km}$ and $36\,\text{km}$ for all waves, and additionally $\lambda_{\text{v}}=18$ for equatorial Rossby waves and Kelvin waves. The vertical wavelength is related to the vertical wavenumber via $m=2\pi/\lambda_{\text{v}}$. The wavelengths selected for plotting in Figures \ref{fig:spec1} were chosen so that the dispersion curves intersected each the Kelvin, Rossby--Kelvin, and equatorial rossby waves shown in Figure \ref{fig:waves1}. There is good agreement between the vertical wavelengths selected, and those that may be inferred from Figure \ref{fig:waves_vert1}.} Equation \eqref{eq:depth} is used to relate the equivalent depth to the vertical wavenumber, which introduces a dependence of $\omega$ on the buoyancy frequency $N$ and pressure scale height $H=RT_{\text{s}}/g$. When computing the dispersion curves, we use $T_{\text{s}}=92\,\text{K}$ to compute $H$, and set $N=6.25\times10^{-3}\,\text{s}^{-1}$, based on Figure \ref{fig:NT}. For a strongly superrotating atmosphere, it is appropriate to set the planetary rotation rate $\Omega$ to that rotating with the equatorial jet \citep{1990JAtS...47..293D,2006JAtS...63.1623I}, so we use $\beta=2\Omega_{\text{eff}}/a=2\overline{u}/a^{2}$, with $\overline{u}=130\,\text{m\,s}^{-1}$. 

Each of the solutions to equations \eqref{eq:sw1}\,--\,\eqref{eq:sw3} have meridional structure functions that decay exponentially away from the equator according to $\hat{X}(y)\propto\exp(-y^{2}/l_{\text{d}}^{2})$, where $l_{\text{d}}$ is called the equatorial deformation radius, defined by \begin{equation}
    l^{2}_{\text{d}}=\frac{2\sqrt{gh_{l}}}{\beta}=\frac{2N}{\beta[m^{2}+1/(4H^{2})]^{\nicefrac{1}{2}}}
\end{equation}
This is the expression used to compute the theoretical deformation radii quoted in the text. We set $N$ and $H$ as stated above, and, also as above, take $\Omega$ to be the rotation rate moving with the equatorial jet. When $l_{\text{d}}$ is quoted as a `meridional extent' (degrees latitude) in the main text, this quantity is computed as $\texttt{meridional extent}=(180^{\circ}/\pi)\times(l_{\text{d}}/a)$.

\bibliography{bibliography}{}
\bibliographystyle{aasjournal}

\end{document}